\documentclass[10pt,aps,prx,twocolumn,superscriptaddress, nolongbibliography]{revtex4-1}

\usepackage{stmaryrd}
\usepackage{epsfig}
\usepackage{subfigure}
\usepackage[colorlinks=true,citecolor=blue,linkcolor=blue]{hyperref}
\usepackage{verbatim}
\usepackage{xcolor}

\usepackage{lmodern}
\usepackage[T1]{fontenc}
\setcitestyle{super}

\usepackage[utf8]{inputenc}
\usepackage{graphicx}
\usepackage{xcolor}
\usepackage{amsmath,amsthm,amssymb,dsfont,mathrsfs}

  \def \a{{\alpha}}
  
  \def \g{{\gamma}}
  \def \dl{{\delta}}
  \def \e{{\epsilon}}

  \def \p{{\pi}}

  \def \ta{{\tau}}

  \def \w{{\omega}}

  \def \G{{\Gamma}}

  \def \vr{{\varrho}}
  
  \def \vk{{\varkappa}}

  \def\hb{\hbar}

  \def\cA{\mathcal{A}}

\def\yd{\text{d}}

  \def \inv{^{-1}} 

\def \vv{{\boldsymbol{v}}}
\def \vk{{\boldsymbol{k}}}
\def \vr{{\boldsymbol{r}}}

\def \vq{{\boldsymbol{q}}}

\begin{document}

\title{Hyperbolic phonon-polariton electroluminescence in graphene-hBN van der Waals heterostructures}

\author{
Qiushi Guo$^{1,2,3*}$, Iliya Esin$^{4*,\dagger}$, Cheng Li$^*$, Chen Chen$^{1}$, Song Liu$^{5}$, James H. Edgar$^{5}$, Selina Zhou$^{6}$, Eugene Demler$^{7}$, Gil Refael$^{4}$, Fengnian Xia$^{1,\dagger}$ \\
\textit{
$^1$Department of Electrical Engineering, Yale University, New Haven 06511, USA.
\\$^2$Photonics Initiative, Advanced Science Research Center, City University of New York, NY, USA\\ 
$^3$Physics Program, Graduate Center, City University of New York, New York, NY, USA\\
$^4$Department of Physics and Institute for Quantum Information and Matter, California Institute of Technology, Pasadena, California 91125, USA\\
$^5$Tim Taylor Department of Chemical Engineering, Kansas State University, Manhattan, Kansas 66506, USA\\
$^6$Department of Electrical Engineering, California Institute of Technology, Pasadena, California 91125, USA\\
$^7$Institute for Theoretical Physics, ETH Zurich, Zürich, Switzerland
$^\ast$These authors contributed equally to this work.\\
$^\dagger$Email: \href{mailto:iesin@caltech.edu}{iesin@caltech.edu}; \href{mailto:fengnian.xia@yale.edu}{fengnian.xia@yale.edu}
}}

\date{\today}

\begin{abstract}
Phonon-polaritons are electromagnetic waves resulting from the coherent coupling of photons with optical phonons in polar dielectrics. Due to their exceptional ability to confine electric fields to deep subwavelength scales with low loss, they are uniquely poised to enable a suite of applications beyond the reach of conventional photonics, such as sub-diffraction imaging and near-field energy transfer. The conventional approach to exciting phonon-polaritons through optical methods, however, necessitates costly mid-infrared and terahertz coherent light sources along with near-field scanning probes, and generally leads to low excitation efficiency due to the substantial momentum mismatch between phonon-polaritons and free-space photons. Here, we demonstrate that under proper conditions, phonon-polaritons can be excited all-electrically by flowing charge carriers. Specifically, in hexagonal boron nitride (hBN)/graphene heterostructures, by electrically driving charge carriers in ultra-high-mobility graphene out of equilibrium, we observe bright electroluminescence of hBN’s hyperbolic phonon-polaritons (HPhPs) at mid-IR frequencies. The HPhP electroluminescence shows a temperature and carrier density dependence distinct from black-body or super-Planckian thermal emission. Moreover, the carrier density dependence of HPhP electroluminescence spectra reveals that HPhP electroluminescence can arise from both inter-band transition and intra-band Cherenkov radiation of charge carriers in graphene. The HPhP electroluminescence offers fundamentally new avenues for realizing electrically-pumped, tunable mid-IR and THz phonon-polariton lasers, and efficient cooling of electronic devices.
\end{abstract}

\maketitle

Phonon-polaritons (PhPs) are electromagnetic waves that arise from the coherent interaction between photons and optical phonons (i.e., quanta of lattice vibrations) in polar dielectrics at mid-infrared (mid-IR) and terahertz (THz) frequencies. They allow for the confinement of electromagnetic fields and energy to length scales significantly smaller than the free photon wavelength, but without incurring significant Ohmic losses often associated with plasmon-polaritons\cite{maier2007plasmonics}. These unique properties make PhPs attractive for various applications, including sub-wavelength waveguiding\cite{dai2014tunable}, near-field sub-diffraction imaging and focusing\cite{li2015hyperbolic}, enhanced near-field heat transport\cite{shen2009surface,song2015enhancement,narayanaswamy2008near}, and controlling the coherence of thermal emission\cite{le1997experimental,greffet2002coherent,yu2023manipulating}. Usually, PhP modes are excited using optical techniques, which not only necessitates costly mid-IR and THz coherent light sources and near-field scanning probes, but also suffers from low excitation efficiency due to the substantial momentum mismatch between PhPs and free-space photons\cite{dias2019fundamental}. Although all-electrical methods to excite plasmons-polaritons\cite{bharadwaj2011electrical} and exciton-polaritons\cite{schneider2013electrically,graf2017electrical,bhattacharya2014room} have been demonstrated, an efficient, all-electric approach to excite and control PhPs still remains elusive. 
\vspace{1.0mm}

\begin{figure*}[ht]
\centering
\includegraphics[width=0.7\linewidth]{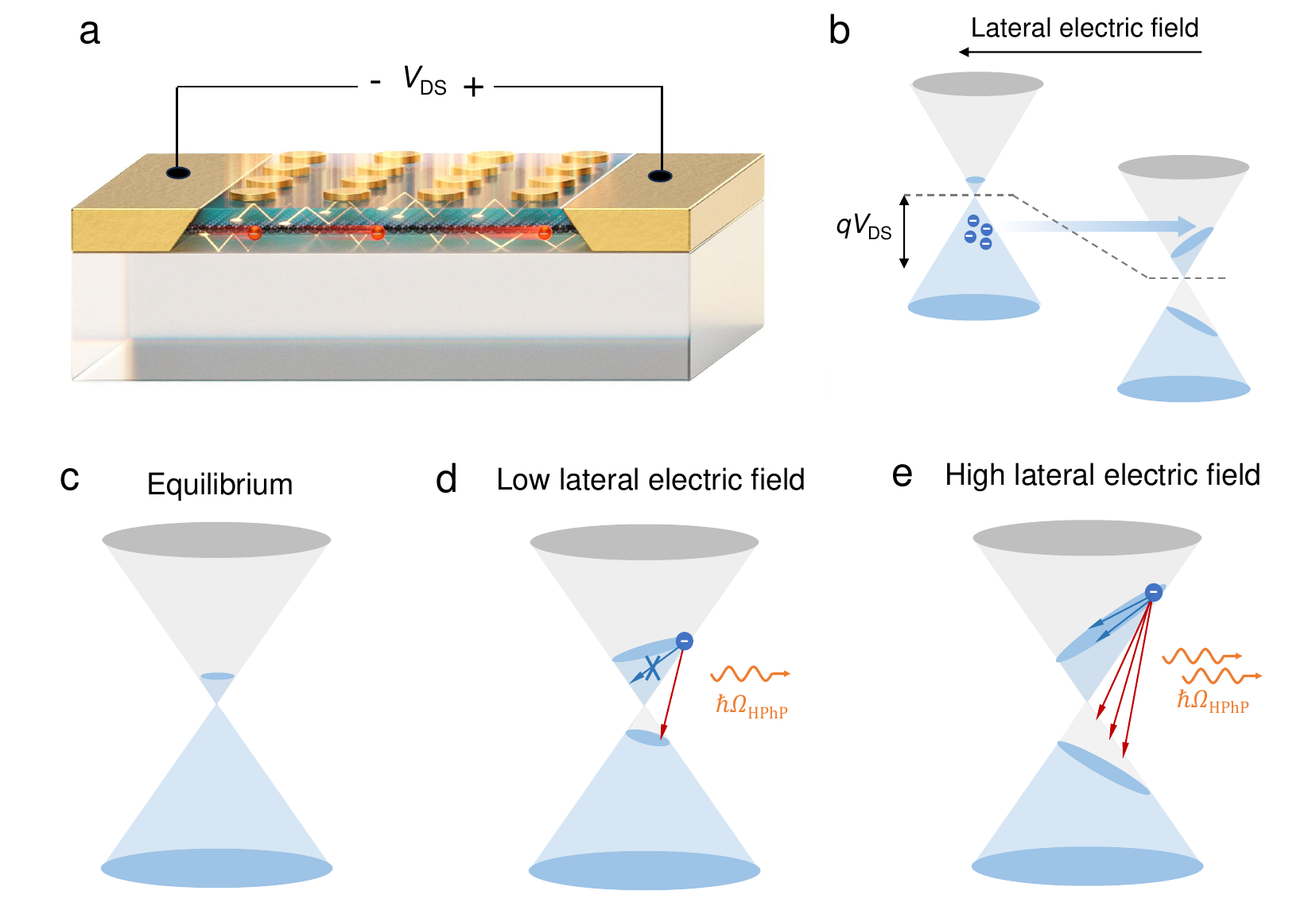}\label{figure1}
\caption{\textbf{Concept of HPhP electroluminescence from non-equilibrium charge carriers in graphene} 
(a) Schematic of HPhP electroluminescence in an hBN/graphene/hBN vdW heterostructure. (b) Illustration of the Zener-Klein tunneling mechanism and Fermi surface tilting. Under a lateral electrical field pointing to the left, electrons in the valence band of graphene can tunnel into the conduction band, generating additional electron-hole pairs in the conduction and valence bands. Meanwhile, the lateral electric field leads to the tilting of graphene’s Fermi surface in momentum space. (c) Graphene charge carrier distribution in equilibrium state. (d) HPhP electroluminescence under a low lateral electric field. (e) HPhP electroluminescence under a high lateral electric field. In both d and e, red arrows denote HPhP emission due to inter-band electron transitions, and blue arrows denote HPhP emission due to intra-band Cherenkov radiations.}\label{Fig1}
\end{figure*}

In fact, electron-phonon interactions are ubiquitous in conducting materials, in which drifting electrons with sufficient kinetic energy can emit optical phonon modes (OPs) in surrounding polar dielectrics via long-range Fr\"ohlich coupling\cite{wang1972electron}. Given that PhPs originate from the hybridization of OPs and photons, this raises an intriguing question: Can PhPs be emitted by drifting electrons? 
In optically isotropic polar dielectrics, such as $\rm SiO_2$ and SiC, photons are unable to effectively hybridize with phonons in the bulk of the material. This limitation arises from the fact that they have comparable energies only in a narrow range of momenta, which is given by the phonon frequency divided by the speed of light. Consequently, PhP modes are confined to the surface, where the photons are allowed to have high in-plane momenta and low frequencies~\cite{caldwell2015low}. 
This constraint has thus far precluded the observation of PhP emission by drifting electrons. Interestingly, such a constrain can be overcome when considering a conducting material surrounded by a hyperbolic polar dielectric, which has in-plane ($\epsilon_{xx}$) and out-of-plane permittivities ($\epsilon_{zz}$) of different signs within the Reststrahlen bands (i.e. $\epsilon_{xx}\epsilon_{zz}<0$). As a distinguishing feature of hyperbolic polar dielectric in the Reststrahlen bands, deep-subwavelength, ray-like electromagnetic modes of arbitrarily large momentum (known as hyperbolic phonon polaritons (HPhPs)) can propagate inside the volume of the dielectric\cite{caldwell2014sub,li2015hyperbolic,giles2016imaging,hu2020topological}. With proper out-coupling schemes, these HPhPs can be scattered into the free space\cite{pons2019launching,castilla2020plasmonic}. Furthermore, these HPhP modes possess a high density of states in momentum space\cite{dai2014tunable,principi2017super}, which greatly enhances their interaction with drifting electrons. 
\vspace{1.0mm}

The graphene/hexagonal boron nitride (hBN) van der Waals (vdW) heterostructure represents an ideal platform for observing the HPhP emission by drifting electrons. First, hBN is a natural hyperbolic polar dielectric, which supports low-loss HPhPs at mid-IR frequencies\cite{low2017polaritons}. Second, the graphene/hBN interface is ultra-clean, yielding reduced impurity scattering and significantly higher carrier mobility and drift velocity in graphene\cite{wang2013one,yamoah2017high}. Third, the electron-acoustic phonon scattering is intrinsically weak in graphene/hBN vdW heterostructure, thus rendering the electron-HPhP scattering dominant ~\cite{principi2017super,Ashida2023}. Prior studies have shown that in hBN-encapsulated graphene, the cooling rate of hot carriers in graphene becomes exceptionally high \cite{yang2018graphene,tielrooij2018out,principi2017super,brasington2022phonon}, suggesting that the presence of hBN's HPhPs can be responsible for the efficient cooling of hot carriers in graphene. Yet, the direct observation of HPhP electroluminescence has not been reported. In this work, we employ mid-IR emission spectroscopy to reveal the hyperbolic phonon-polariton electroluminescence within graphene/hBN van der Waals heterostructures.

\begin{figure*}[ht]
\centering
\includegraphics[width=0.8\linewidth]{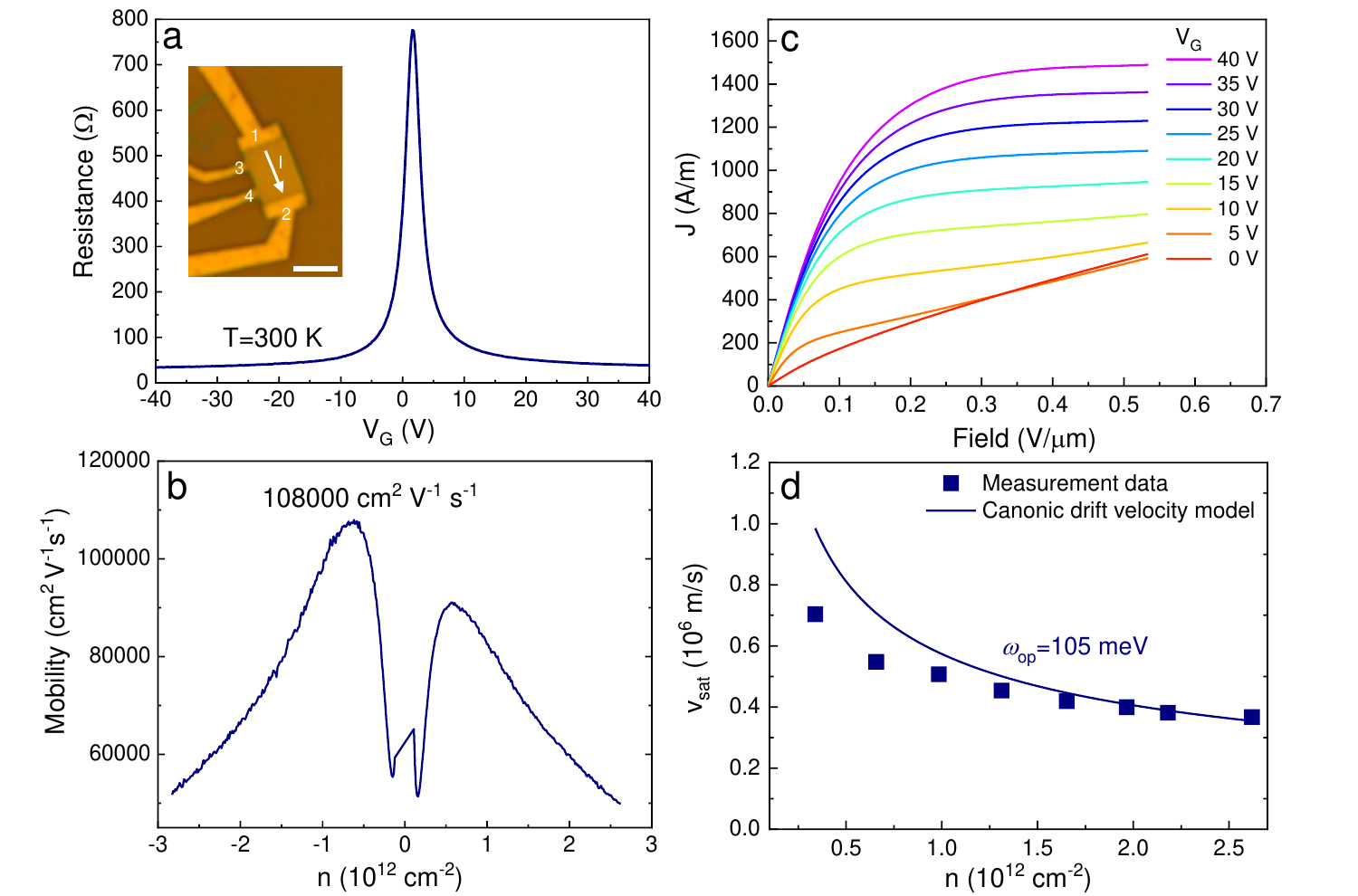}
\caption{\textbf{Room-temperature charge carrier transport in ultra-clean graphene encapsulated by h$^{10}$BN} (a) Gate-dependent graphene resistivity measured at room temperature. The inset shows the optical image of the experimental device with a hall-bar geometry, which has a channel width of 5 $\mu$m. (b) Extracted room-temperature carrier mobility of $^{10}$B hBN encapsulated graphene as a function of charge carrier density. (c) Measured room-temperature current density ($J$) versus lateral electric field ($F$) at various gate voltages ($V_\mathrm{G}$). (c) Extracted saturation velocity ($v_\mathrm{sat}$) versus carrier density. Blue symbols are experimental data. The solid blue curve is an estimation of $v_\mathrm{sat}$ by a canonic drift velocity model considering the lowest energy optical phonon mode of h$^{10}$BN, which has an energy of $\hbar\omega_\mathrm{OP}=105$ meV.}\label{Fig2}
\end{figure*}

Figure \ref{Fig1}a depicts the concept of HPhP electroluminescence in an hBN/graphene/hBN vdW heterostructure. In this setup, a single-layer
of graphene sheet is encapsulated by two hBN slabs located below and above the graphene sheet. When a lateral electric field is applied along the graphene channel, charge carriers in graphene (red spheres) are accelerated. Under certain conditions, they can efficiently dissipate their kinetic energy by emitting HPhP modes (bright yellow rays) in nearby hBN slabs. These HPhP modes in hBN slabs, which possess large momentum, can be subsequently out-coupled to the free space via the nano-scatters on top of the hBN. As illustrated in Fig. \ref{Fig1}b, the effect of the lateral electric field is two-fold. First, it generates additional electron-hole pairs in graphene through the Zener-Klein tunneling mechanism\cite{vandecasteele2010current,yang2018graphene,schmitt2023mesoscopic} with the rate proportional to $\sim F^{3/2}$ \cite{berdyugin2022out}. Second, it leads to an energy difference between the forward populated and unpopulated charge carrier states (manifested by the tilting of graphene's Fermi surface in momentum space\cite{andersen2019electron}), providing a finite charge carrier drift velocity $v_{\yd}$.  
\vspace{1.0mm}

\begin{figure*}[ht]
\centering
\includegraphics[width=0.8\linewidth]{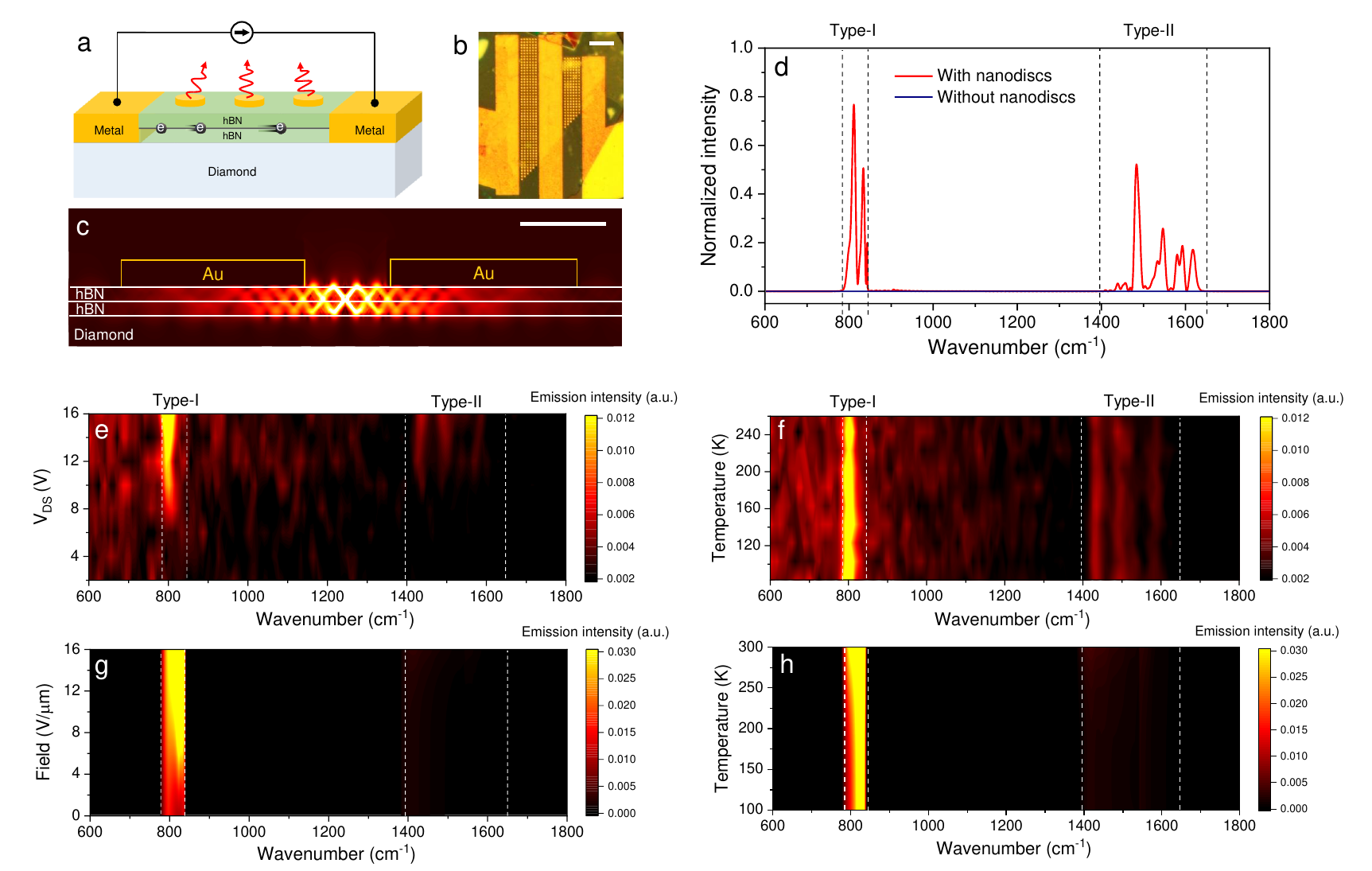}
\caption{\textbf{Mid-IR HPhP electroluminescence in h$^{10}$BN/graphene/h$^{10}$BN heterostructure.} (a) Illustration of HPhP electroluminescence. The experimental device consists of a h$^{10}$BN/graphene/h$^{10}$BN heterostructure fabricated on a diamond substrate. Edge contacts are formed between the graphene channel and Au/Cr metal leads. Au nanodiscs (diameter = 700 nm, height = 100 nm, pitch= 1 $\mu$m) on top of hBN layer enable out-coupling of HPhPs to free space. (b) Optical microscope image of the fabricated device. Scale bar: 10 $\mu$m. (c) Simulated near-field distribution of a HPhP mode at 1400 cm$^{-1}$ excited by a dipole emitter placed within the graphene layer. The sharp edges of Au nanodiscs scatter the deep-subwavelength, ray-like HPhP mode into the free space. Scale bar: 300 $\mu$m. (d) Simulated far-field emission spectra with (red) and without (blue) Au nanodiscs on the top of the heterostructure. Measured (e) and calculated (g) $V_\mathrm{DS}$ dependence of HPhP electroluminescent spectrum at $V_\mathrm{G}$=0. Measured (f) and calculated (h) temperature dependence of HPhP electroluminescent spectrum under $V_\mathrm{DS}=16$ V and $V_\mathrm{G}$=0. White dashed lines in e, f, g and h mark the boundaries of type I ($785 - 845$ cm$^{-1}$) and type II ($1394.5 - 1650$ cm$^{-1}$) Reststrahlen bands of h$^{10}$BN.}\label{Fig3}
\end{figure*}

The HPhP electroluminescence under the lateral electric field can be described as a spontaneous emission process of HPhP modes by charge carriers in graphene. This process adheres to the principles of both energy and momentum conservation, with the emission rate determined by Fermi's golden rule. Figure \ref{Fig1}d illustrates the HPhP electroluminescence under a low lateral electric field. Compared to the equilibrium state (Fig. \ref{Fig1}c), the Zener-Klein tunneling mechanism generates additional electrons and holes in the conduction band and valence band. Due to the additional holes in the valence band, electrons in the conduction band can undergo a transition to the valence band, leading to the emission of HPhP modes via inter-band transition (red arrow). However, since the energy difference between forward and backward-populated electrons is less than the energy of HPhPs, drifting electrons in graphene cannot emit HPhP modes through an intra-band process (blue arrow) because there are no available final states for such transitions to occur. As shown in Fig. \ref{Fig1}d, under a sufficiently high lateral electric field, there will be more electron-hole pairs generated. Meanwhile, the energy difference between forward and backward-populated electrons can exceed the energy of HPhPs. In this case, the HPhPs electroluminescence can be more efficient since charge carriers can emit HPhPs through both inter-band transitions and intra-band transitions, as shown in Fig.~\ref{Fig1}e. The intra-band emission is analogous to phonon Cherenkov emission, originally describing a mechanism of acoustic phonon emission when the drift velocity exceeds the speed of sound~\cite{Harold1962,Pippard1963,Komirenko2000,Huang2005,Suresha2006,Andersen2019}. However, the HPhP emission rate is much higher than the conventional phonon-assisted electron-hole scattering,  due to the larger density of states available for an HPhP emission.


To validate the strong interactions between charge carriers in graphene and HPhPs, we first measure the electrical transport characteristics of a hBN-encapsulated graphene device fabricated by a vdW assembly technique\cite{wang2013one}. Here, we used $^{10}$B isotopically enriched hBN (h$^{10}$BN)\cite{liu2018single} for graphene encapsulation. Compared to naturally abundant hBN, h$^{10}$BN offers around threefold improvement in HPhP lifetime\cite{giles2018ultralow}. In our device, the thickness of the SiO$_2$ substrate, the bottom hBN, and the top hBN are 300 nm, 35 nm, and 42 nm, respectively.  To eliminate the effect of metal/graphene contact resistance, we employed a four-terminal (Hall-bar) device structure (inset of Fig. \ref{Fig2}a) to measure charge carrier transport in the graphene. Figure \ref{Fig2}a shows the measured room-temperature resistivity of the device as a function of gate voltage ($V_\mathrm{G}$) at a low bias voltage of 10 mV. The transport characteristics indicate the graphene is remarkably pristine, yielding a room-temperature carrier mobility ($\mu_\mathrm{e}$) exceeding 100,000 cm$^2$ V$^{-1}$ s$^{-1}$ at a carrier density of $\sim 1\times10^{12}$ cm$^{-2}$ (Fig. 2b), which is close to the acoustic phonon-limited mobility for intrinsic graphene\cite{Hwang2008}. 
\vspace{1.0mm}

Using the same device, we further measured the current density ($J$) as a function of lateral electric field ($F$) at room temperature. As shown in Fig. \ref{Fig2}c, under a low lateral electric field ($F<0.05$ V/$\mu$m), the current density scales almost linearly with the electric field. In this regime, the carrier drift velocity is mainly limited by the electron-acoustic phonon scattering~\cite{Hwang2008}. However, as the lateral electric field strength increases, the device exhibits current density saturation~\cite{Meric2008} since the carrier drift velocity $v_{\text{d}}$ reaches its maximum achievable value, determined by the group velocity of graphene's Dirac cone ($v_\text{F} = 10^6$ m/s)~\cite{Berdyugin2022}. As shown in Fig. \ref{Fig2}c, at even higher lateral electric fields, the current further saturates. This phenomenon occurs because the forward and backward-populated charge carriers can accumulate a sufficient energy difference. Such an energy difference enables the forward-populated carriers to dissipate their kinetic energy by emitting optical phonons, leading to a further reduction of drift velocity (see Supplementary Information Section I for detailed analysis). In Fig. \ref{Fig2}d, we plot the saturation velocities ($v_\mathrm{sat}$) (blue symbols ) of graphene at various carrier concentrations. By comparing the measured $v_\mathrm{sat}$ with a canonical drift velocity model\cite{yamoah2017high} (blue solid curve), we found that the $v_\mathrm{sat}$ of graphene is primarily limited by the remote OP modes of h$^{10}$BN ($\omega_\mathrm{OP}$= 105 meV). This finding strongly suggests that, under high lateral electric fields, charge carriers in graphene interact strongly with OP modes in hBN, raising the intriguing possibility of observing HPhP electroluminescence.


\vspace{1.1mm}

Next, we study the HPhP electroluminescence in h$^{10}$BN/Graphene/h$^{10}$BN heterostructures using mid-IR emission spectroscopy (see Supplementary Information Section II for details) and an experimental device illustrated in Fig. \ref{Fig3}a. Figure \ref{Fig3}b shows an optical image of our fabricated device. We fabricate the device on a diamond substrate since the diamond does not support any optical phonon modes within the frequency range of interest. In addition, due to its exceptionally high thermal conductivity, diamond can efficiently dissipate the lattice heat generated in graphene and hBN, effectively minimizing the influence of thermal emission in our experimental observations. Compared to the device we used for the transport measurements (inset of Fig. \ref{Fig2}a), the device has a larger graphene area of $\sim$140 $\mu$m$^2$, with a graphene channel length of 7 $\mu$m. This augmented active region ensures sufficient emission power and a substantial signal-to-noise ratio. In addition, as shown in Fig. 3a and b, we fabricated an Au nanodiscs array (diameter = 700 nm, height = 100 nm, pitch = 1 $\mu$m) on top of the hBN layer in order to significantly enhance the out-coupling of HPhPs into free space. As shown in the near-field electromagnetic simulation (Fig. \ref{Fig3}c), the sharp bottom edges of Au nanodiscs can compensate for the large momentum mismatch between the HPhP modes and free-space electromagnetic waves, thus scattering the propagating HPhP mode in hBN into free space\cite{dai2017efficiency,pons2019launching}. Moreover, the simulated emission spectra shown in Fig. \ref{Fig3}d further verify that our designed nanodiscs array is capable of scattering the HPhPs into the far field. 
\vspace{1.0mm}

Figure \ref{Fig3}e shows the measured electroluminescence spectrum as a function of $V_\mathrm{DS}$ at zero $V_\mathrm{G}$ and an environmental temperature of 83 K. Notably, broadband emission features in type I Reststrahlen band can be detected at a modest $V_\mathrm{DS}$ of 6 V, which corresponds to a lateral electric field ($F$) of $\sim$0.9 V/$\mu$m and an electrical power of $\sim$20 mW. Furthermore, as $V_\mathrm{DS}$ exceeds 9 V, broadband emission features within the type-II Reststrahlen bands are detectable. In addition, the intensity of electroluminescence in the type I Reststrahlen band is always stronger than that in the type II Reststrahlen band. Remarkably, as shown in Fig. \ref{Fig3}f, under an applied $V_\mathrm{DS}$ of 16 V, the intensity of electroluminescence remains nearly unchanged even as we elevate the environmental temperature from 83 K to 273 K. This phenomenon strongly suggests that the observed electroluminescence cannot be attributed to traditional black-body or Super-Planckian thermal radiation process, since they typically exhibit a strong dependence on the temperature of the emitter\cite{principi2017super,guo2012broadband}. To substantiate this point, we also investigated the traditional thermal radiation process by conducting a control experiment using another hBN-encapsulated graphene device without Au nanodiscs to out-couple HPhPs. With this device, the emission features emerge at a significantly higher applied electrical power ($>500$ mW), thus showing very low emission efficiency. More importantly, the emission intensity dramatically decreases at lower environmental temperatures, which is in stark contrast to our observation here (see Supplementary Information Section III for details).  
\vspace{1.0mm}

\begin{figure*}[ht]
\centering
\includegraphics[width=0.9\linewidth]{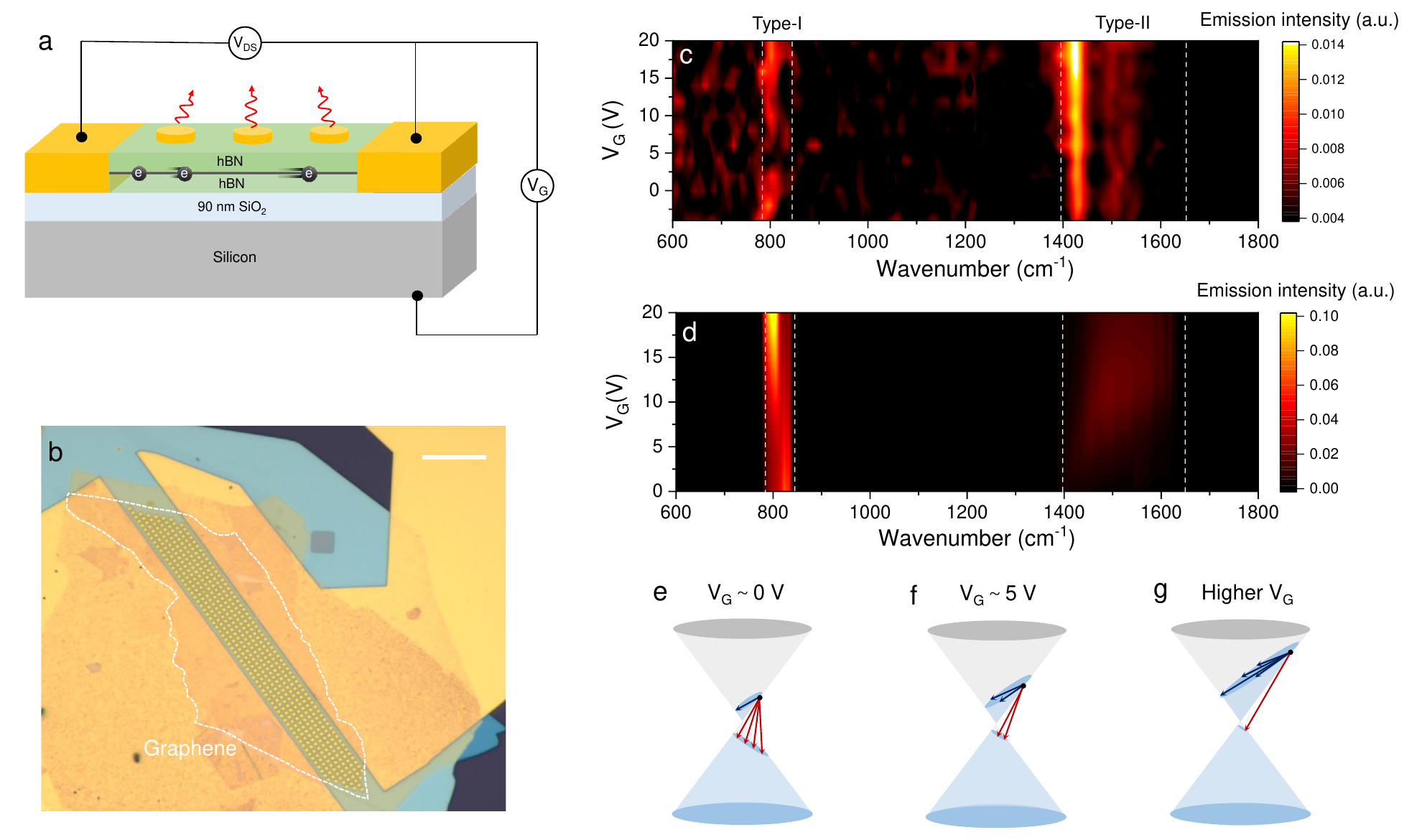}
\caption{\textbf{Gate-voltage dependence of HPhP electroluminescence} (a) Illustration of the experimental device in which h$^{10}$BN/graphene/h$^{10}$BN heterostructure was fabricated on SiO$_2$ (90 nm)/silicon substrate. The doping concentration of graphene is controlled by the back-gate voltage applied to the silicon substrate. (b) Optical image of the experimental device for gate-voltage dependent HPhP emission measurements. Scale bar: 10 $\mu$m. (c) Measured gate-voltage dependence of HPhP electroluminescence spectrum at 100 K. In the measurement, the $V_\mathrm{DS}$ was fixed at 16 V. (d) Calculated gate-voltage dependence of e-HPhP scattering rate at a $V_\mathrm{DS}$ of 16 V. (e) Illustration of HPhP electroluminescence at a low $V_\mathrm{G}$ (left), an intermediate $V_\mathrm{G}$ (middle), and a high $V_\mathrm{G}$ (right).  }\label{Fig4}
\end{figure*}

To gain deeper insights into these behaviors, we model the emission by kinetic equation~\cite{Andersen2019} for the HPhPs occupations $b_{\vq, a}(\vr,t)$, where $a=(\a,n)$ with $\a=I,II$ denoting the Restrahlen band, $n=0,1,2...$ denoting the branch of HPhP modes, and $\vq$ is the wavevector.
The kinetic equation can be written as
\begin{equation}
    \frac{\partial b_{\vq, a}}{\partial t}=\G^{\rm amp}_{\vq,a} b_{\vq, a}+\g^{\rm em}_{\vq, a}-\frac{b_{\vq, a}-b^0_{\vq, a}}{\ta_p}-\vv_{s,\vq, a} \cdot \nabla b_{\vq,a}
\label{eq:PhononicKinetic}
\end{equation}
where $\Gamma_{\vq,a}^{\rm amp}$ and $\g_{\vq, a}^{\rm em}$ denote the amplification and spontaneous emission rates of HPhP modes due to the emission and absorption of HPhPs by the non-equilibrium electron and hole populations in graphene. $\vv_{s,\vq, a}=\partial_\vk \w_{\vq, a}$ is the group velocity of the HPhP modes, and $\w_{\vq,a}$ is the angular frequency of HPhP modes, $b_{\vq, a}^0=(e^{\hbar\w_{\vq, a}/k_BT}-1)\inv$ is the equilibrium distribution of HPhPs. $1/\tau_p$ is the loss rate of HPhP considering phonon-phonon interactions, phonon-impurity interactions, and out-coupling to free space. The detailed calculations of  $\Gamma_{\vq,a}^{\rm amp}$ and $\gamma_{\vq,a}^{\rm em}$ are described in Supplementary Information Sections IV to VI. In our theoretical calculation, we accounted for all energy and momentum allowed e-HPhP scattering processes, given the non-equilibrium electronic distribution. The average emission rate per frequency $\w$ in the steady state is then given by 
\begin{equation}
I_{\rm rad}(\w)=\int \frac{d^2\vq}{(2\p)^2}\sum_{a}\frac{\hb\w_{\vq, a}}{\ta_{\rm rad}}\bar b_{\vq,a}\dl(\w-\w_{\vq,a }),\label{eq:rate}
\end{equation}
where $\tau_{\rm rad}$ denotes the coupling of the HPhPs modes to free space assisted by the Au nanodiscs, and $\bar b_{\vq,a}=\cA^{\inv}\int d^2\vr b_{\vq,a}(\vr)$ is the HPhP occupation averaged over the sample area $\cA$.
\vspace{1.0mm}

Figures \ref{Fig3}g and h show the $V_\mathrm{DS}$ dependence and temperature dependence of averaged HPhP emission rate calculated by equation \ref{eq:rate}. The theoretical results agree with our experiments, indicating that the e-HPhP scattering process is responsible for the observed electroluminescence. The higher observed emission intensity in the type I Reststrahlen band can be attributed to the lower energy threshold needed to excite this type of HPhP mode. Moreover, the emission intensity does not change significantly with temperature, because the temperature mostly affects the mobility of the carriers~\cite{Hwang2008}, which has a weak effect on the e-HPhP scattering rate. Furthermore, according to our theoretical analysis, we did not observe a net gain of HPhP along the graphene channel. This is presumably due to the level of the disorder in graphene and the strong out-coupling of HPhP modes to the free space. 
\vspace{1.0mm}

As HPhP electroluminescence arises from electron-HPhP scattering, its intensity can be significantly influenced by the Fermi level of graphene. This is because the Fermi surface of graphene governs the available momentum space for e-HPhP scattering events. To investigate this, we experimentally study $V_\mathrm{G}$ of the HPhP electroluminescence using another device illustrated in Fig. \ref{Fig4}a. For this device, the h$^{10}$BN/Graphene/h$^{10}$BN heterostructure was fabricated on SiO$_2$ (90 nm)/silicon substrate and the Fermi level of graphene can be controlled by the back-gate voltage ($V_\mathrm{G}$) applied on the silicon substrate. Figure \ref{Fig4}b shows the optical image of the fabricated device, which has a graphene channel length of 7 $\mu$m.
\vspace{1.0mm}

Fig. \ref{Fig4}c shows the measurement results under a fixed lateral bias voltage of 16 V. Again, we observed broadband HPhP emission features in both type-I and type-II Reststrahlen bands, despite that the observed emission intensity in type-I Reststrahlen is weaker than that of type-II in this case. This can be ascribed to the high optical loss of SiO$_2$ substrate around 800 cm$^{-1}$ due to the presence of an optical phonon mode\cite{yan2013damping}. Interestingly, we found the HPhP emission in type-I and type-II Reststrahlen bands behave quite differently in response to the gate voltage ($V_\mathrm{G})$. As  $V_\mathrm{G}$ was tuned from -2 V to 22 V, the intensity of type-I HPhP emission first decreases, and then substantially increases. However, the type-II HPhP emission monotonously increases with the gate voltage. This behavior, which is well captured by the theoretically calculated HPhP emission rate shown in Fig. \ref{Fig4}d, is attributed to the different responses of inter-band and intra-band emission processes to the gate voltage. Specifically, when $V_\mathrm{G}$ $\sim$ 0, the system predominantly emits type-I HPhPs through the inter-band processes, whereas the intra-band emission process is weak due to the small Fermi surface and momentum space, as illustrated in \ref{Fig4}e.  As $V_\mathrm{G}$ increases, the density of holes decreases, leading to a suppression of the inter-band emission process and a strong reduction of type-I HPhPs emission. Further increasing the $V_\mathrm{G}$ resulting in stronger intra-band emission of type-I HPhPs due to larger available momentum space. In contrast, type-II HPhPs are predominantly emitted through inter-band processes within the parameter range of the experiment. Due to the higher energy of type-II HPhPs, their emission is activated only for higher values of $V_{\rm G}$, giving rise to a growth of emission with the $V_{\rm G}$. For even higher values of $V_{\rm G}$, the type-II HPhP emission rate starts to decrease because of the decreasing density of holes, similar to type-I HPhPs. 
\vspace{1.0mm}

To summarize, our study presents the first observation of HPhP electroluminescence in graphene/hBN van der Waals heterostructures, which also marks the first experimental demonstration of exciting phonon-polariton waves exclusively through electrical methods. This observed HPhP electroluminescence emerges under a strongly biased non-equilibrium electronic state in graphene, which supports both intra-band Cherenkov and inter-band Klein-Zener emission processes. Notably, HPhP electroluminescence exhibits a broadband emission spectrum within the two Reststrahlen bands of hBN, characterized by temperature independence that distinguishes it from traditional black-body and Super-Planckian thermal radiation. Moreover, our study further reveals that the charge carrier density in graphene can modulate the respective contributions of intra-band Cherenkov and inter-band Klein-Zener emission processes to the HPhP electroluminescence. 
\vspace{1.0mm}

Although we have not observed a net gain of HPhP along the device channel (see Supplementary Information Section VI for the theoretical efficiency analysis) in this study,  achieving a net gain holds significant promise by minimizing HPhP outcoupling to free space and further increasing the driving current. 
Furthermore, by introducing the feedback of HPhP modes using polaritonic nano-resonators\cite{caldwell2014sub,tamagnone2018ultra,brown2018nanoscale}, the HPhP electroluminescence can potentially become coherent, thus leading to the electrically driven HPhP lasing. Looking forward, our reported HPhP electroluminescence mechanism can be harnessed to develop conceptionally new and power-efficient electrically-pumped amplifiers, and polaritonic lasers at mid-IR and THz frequencies for sensing and communication applications. On a different front, the HPhP electroluminescence or lasing can serve as an efficient mechanism for cooling hot carriers in silicon or III-V semiconductor-based CMOS, RF, and power transistors, which can greatly alleviate the thermal bottleneck of current CMOS, RF, and power electronics.


\section*{Methods}
\textbf{Device Fabrication.} h$^{10}$BN crystals are synthesizes using 
 a method described in Ref. \cite{liu2018single}. Graphene and h$^{10}$BN flakes were obtained through mechanical exfoliation. The process of device fabrication began by assembling a h$^{10}$BN/graphene/h$^{10}$BN heterostructure on a 90 nm SiO$_2$/Si (or diamond) substrate, following the method described in Ref. \cite{wang2013one}. Subsequently, the sample underwent annealing at 600 $^\circ$C for 6 hours in a flowing hydrogen/argon environment. This annealing step served to eliminate impurities and enhance the carrier mobility of graphene. To define the device channel and metal electrodes, a 100 kV electron beam lithography system and PMMA as a resist were employed. The heterostructure was partially etched using SF$_6$ and O$_2$ plasma in an Oxford Plasmalab 100 reactive ion etching (RIE) system. Following this, Cr/Au electrodes (5 nm/45 nm) were thermally evaporated to form edge contacts with the graphene. To fabricate the Au nanodiscs on top of the heterostructure, another e-beam lithography step was used to define the pattern. Subsequently, we etched the top h$^{10}$BN by $\sim$4 nm by CH$_3$ and O$_2$ plasma. This step resulted in the formation of h$^{10}$BN nano-slots, which stabilized the Au nanodiscs. Finally, the Au discs were fabricated by evaporating 5 nm Cr and 100 nm Au using an e-beam evaporator, followed by a lift-off process in acetone.

\textbf{Mid-infrared electroluminescence measurement.} The detailed mid-infrared electroluminescence measurement procedure is described in Supplementary Information Section II. In the measurement, the sample was wired-bonded to a chip carrier and was placed on a temperature-controllable stage (Linkham HFS600-PB4). The gate voltage to the device was applied to the device using a Keithley 2612 source meter. In this setup, instead of employing a DC source-drain bias voltage, a 34.5 kHz square wave with phase synchronization to the lock-in amplifier was utilized as the source-drain voltage bias. The square wave was generated by an arbitrary function generator (AFG) and amplified by a custom-made voltage amplifier. Once the square wave was applied to the device, the MCT detector collected both a DC environmental thermal emission background signal and an AC emission signal from the device. The AC voltage output from the MCT detector underwent filtering through a narrow pass-band band-pass filter (34.2 - 34.8 kHz). The filtered output voltage signal was then fed into the lock-in amplifier with a time constant set to 1 ms. The DC output signal from the lock-in amplifier was directed to the DC input of the FTIR (Fourier-transform infrared spectroscopy). The FTIR was used to form the interferogram and acquire the IR emission spectra after performing the Fourier transform on the data. 

\textbf{Electromagnetic simulations.} The optical response of graphene-hBN heterostructures was simulated using Ansys Lumerical FDTD (2023b). A 3-D FDTD model was constructed by specifying the thickness and geometry of hBN slabs, graphene, and Au discs. Periodic boundary conditions were imposed on two sides and absorption boundary conditions (PML) were imposed both on the top and the bottom of the simulation region. A graphene layer was modeled using a
uniaxial anisotropic permittivity assuming that the graphene layer has a finite thickness (C (graphene)-Falkovsky (mid-IR) material model). In the model, graphene’s conductivity is described by a Drude-like expression\cite{falkovsky2008optical} which is valid for long-wavelength mid-IR. The the uniaxial anisotropic dielectric function of h$^{10}$BN is modeled by
\begin{equation}
\epsilon_\mu(\omega) = \epsilon_{\infty,\mu}\left( {\frac{{\omega _\mathrm{LO,  
 \mu}^2 - \omega^2 - i\Gamma_\mu \omega }}{{\omega _\mathrm{TO, \mu}^2 - \omega^2 - i\Gamma_\mu \omega }}} \right)\label{hBN dielectric}
\end{equation}

\noindent where $\mu$ stands for in-plane (xy) or out-of-plane (z) direction. $\omega$, $\omega_\mathrm{TO}$, $\omega_\mathrm{LO}$, $\Gamma$ and $\epsilon_\infty$ represent the optical frequency, the TO and LO phonon frequencies, the phonon damping, and the high-frequency permittivity, respectively. According to Ref. \cite{giles2018ultralow}, for  h$^{10}$BN, $\epsilon_{\infty,xy}$=5.1, $\epsilon_{\infty,z}$=2.5, $\omega _\mathrm{LO, xy}$=1650 cm$^{-1}$, $\omega_\mathrm{LO, z}$=845 cm$^{-1}$, $\omega_\mathrm{TO, xy}$=1394.5 cm$^{-1}$, $\omega_\mathrm{TO, z}$=785 cm$^{-1}$, $\Gamma_{xy}$=1.8 cm$^{-1}$, $\Gamma_z$=1 cm$^{-1}$. To model and simulate the excitation and far-field emission of HPhPs (Fig. 3b and c), an electric dipole emitter located within the graphene layer was used as the source. The emitter's bandwidth was chosen to range from 500 to 2000 cm$^{-1}$, and its orientation was set to be in the plane of graphene in order to emulate the electron-HPhP scattering. In order to reduce the simulation time consumption, we set the single graphene layer to be 1 nm in thickness and the corresponding permittivity of graphene was normalized to match the thickness of 1 nm. The mesh size of the simulation was set to 1 nm.


\section*{Data Availability}
The data that support the plots within this paper and other findings of this study
are available from the corresponding author upon reasonable request.

\section*{Code Availability}
The computer codes used to perform the theoretical analysis and simulations in this paper are available from the corresponding author upon reasonable request.

\section*{Acknowledgements}
The authors would like to thank Renwen Yu, Javier Garcia de Abajo, Jacob Khurgin, and Francisco Guinea for insightful discussions. Research at Yale University was supported by the National Science Foundation CAREER Award (ECCS-1552461) and Yale University. E.D. acknowledges support from the SNSF project 200021-212899 and the ARO grant number W911NF-21-1-0184. J.H.E. acknowledges the support of the Materials Engineering and Processing program of the National Science Foundation, Award Number CMMI 1538127.  G.R. expresses gratitude for the support by the Simons Foundation, and the ARO MURI Grant No. W911NF-16-1-0361. I.E. is grateful for support from the Simons Foundation and the Institute of Quantum Information and Matter.

\section*{Authors Contributions}
Q.G. conceived the idea. Q.G. and C.L. fabricated the devices and performed the measurements with assistance from C.C. I.E. performed the theoretical modeling and physical interpretation of the carrier transport and the HPhP electroluminescence, with inputs from G.R. and E.D. S.L. and J.H.E. synthesized the h$^{10}$BN.  Q.G. performed the optical simulation with assistance from S.Z. Q.G. and I.E. wrote the manuscript with inputs from all authors. E.D., G.R., and F.X. supervised the project.

\section*{Competing Interests}
The authors declare no competing interests.

\newpage


%

\end{document}


\title{Supplementary Information for ``Hyperbolic phonon-polariton electroluminescence in graphene-hBN van der Waals heterostructures'' }

\author{
Qiushi Guo$^{1,2,3*}$, Iliya Esin$^{4*,\dagger}$, Cheng Li$^*$, Chen Chen$^{1}$, Song Liu$^{5}$, James H. Edgar$^{5}$, Selina Zhou$^{6}$, Eugene Demler$^{7}$, Gil Refael$^{4}$, Fengnian Xia$^{1,\dagger}$ \\
\textit{
$^1$Department of Electrical Engineering, Yale University, New Haven 06511, USA. \\
$^2$Photonics Initiative, Advanced Science Research Center, City University of New York, NY, USA\\ 
$^3$Physics Program, Graduate Center, City University of New York, New York, NY, USA\\
$^4$Department of Physics and Institute for Quantum Information and Matter, California Institute of Technology, Pasadena, California 91125, USA\\
$^5$Department of Chemical Engineering, Kansas State University, Manhattan, Kansas 66506, USA\\
$^6$Department of Electrical Engineering, California Institute of Technology, Pasadena, California 91125, USA\\
$^7$Institute for Theoretical Physics, ETH Zurich, Zürich, Switzerland\\
$^\ast$These authors contributed equally to this work.}\\
$^\dagger$Email: \href{mailto:iesin@caltech.edu}{iesin@caltech.edu}; \href{mailto:fengnian.xia@yale.edu}{fengnian.xia@yale.edu}
}

\maketitle


\section{Electronic transport in graphene}
In this section, we analyze the electronic distribution and transport properties of graphene under a strong lateral bias field, $F$. We focus on the population near one of the Dirac cones with energy $E_\yc(\vk)=v_\yF|\vk|$ for the conduction band and $E_\yv(\vk)=-v_\yF|\vk|$ for the valence band. We approximate the electron (hole) steady-state distributions, $f_\ye$ ($f_\yh$) in the conduction (valence) band by
\Eq{
f_{\ye(\yh)}(\vk)=\frac{1}{1+e^{\beta[E_{\yc(\yv)}(\vk-\Delta\vk_{\ye(\yh)})-\ve_{\ye(\yh)}]}}
\label{eq:Occupation}
}
Here, $\ve_{\ye(\yh)}$ sets the total density of conduction electrons (valence holes), given by $n_{\ye(\yh)}=\n\int \frac{d^2\vk}{(2\p)^2}f_{\ye(\yh)}(\vk)$ where $\n=4$ is the spin-valley degeneracy of graphene, $\Delta\vk_{\ye(\yh)}$ sets the center-of-mass velocity, and $\beta=1/T$ is the inverse temperature, assumed to be the same for the electron and hole populations. \vspace{1.2mm}

The electron-hole pairs are generated by thermal excitations with the rate $R_\yT$ and by Klein-Zener tunneling. The Zener-Klein tunneling arises from the electron-hole pair creation by the electric-field, known as the Schwinger process. The rate of such electron-hole pair creation scales as $\sim F^{3/2}$~\cite{Allor2008} with a dimensional field-independent prefactor $A_{\rm ZK}$, which we use as a fitting parameter~\cite{Berdyugin2022}. 
In turn, the electron-hole pairs can recombine with a characteristic coefficient $B_\yR$, giving rise to the rate equations for the electron and hole densities:
\begin{subequations}
\begin{eqnarray}
&&\dot n_\ye=A_{\rm ZK} F^{3/2}+R_\yT-B_Rn_\ye n_\yh  \\
&&\dot n_\yh=A_{\rm ZK} F^{3/2}+R_\yT-B_Rn_\ye n_\yh.       
\end{eqnarray}
\label{eq:DensityRateEq} 
\end{subequations}
Here, $B_\yR=B_{\rm nr}+B_{\rm ep}$, where $B_{\rm nr}$ denotes the non-radiative recombination rate primarily due to electron-impurity and electron-electron (Auger) processes and $B_{\rm ep}$ arises due to phonon emission. Tab.~\ref{tab:Parameters} shows the values of the parameters taken in the simulation.
\vspace{1.2mm}

\begin{figure}
  \centering
  \includegraphics[width=13.6cm]{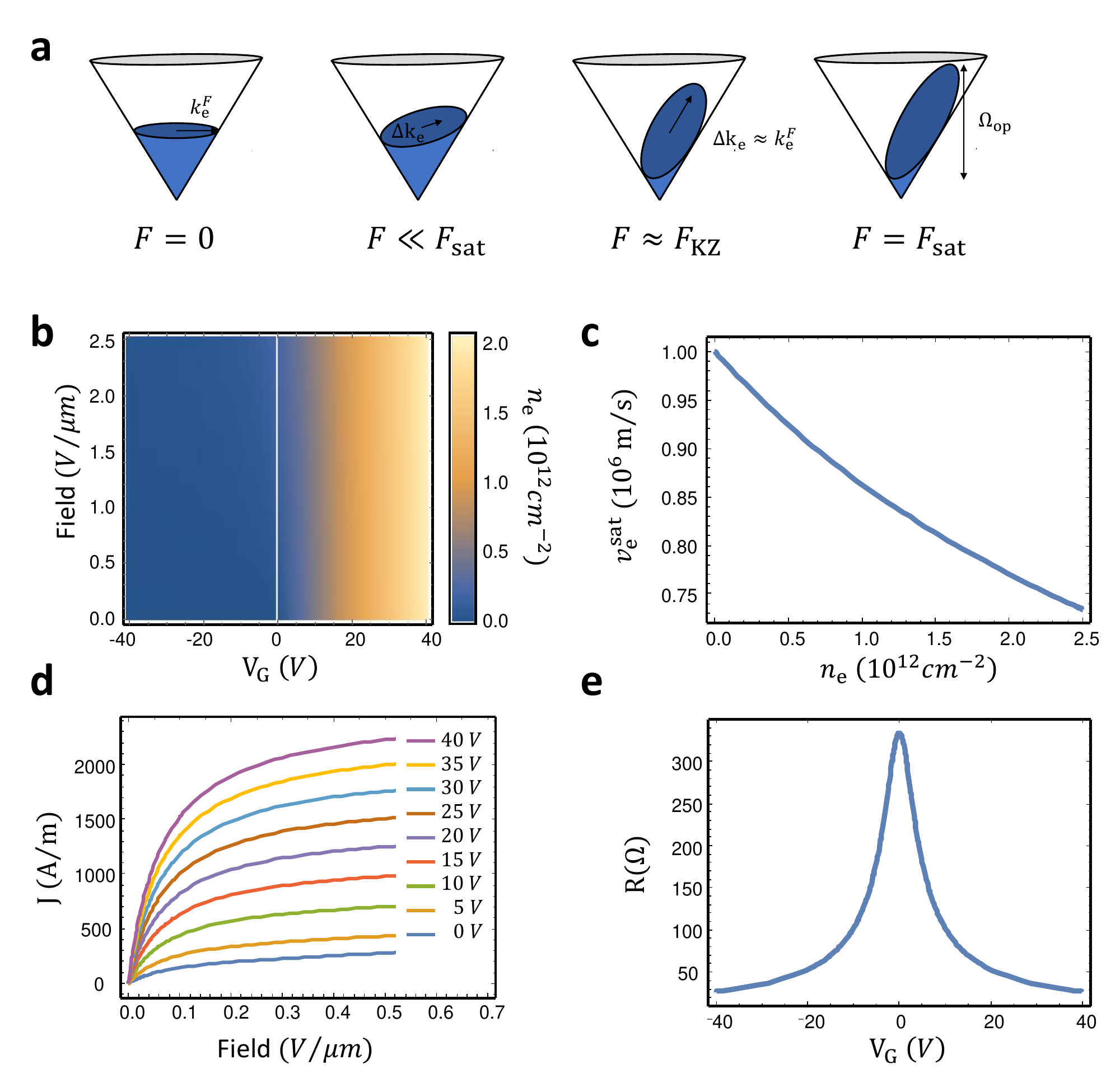}\\
  \caption{ (a) Distribution of the electrons in the conduction band as a function of the electric field, $F$, in different saturation regimes. 
  (b) Electronic density on the conduction band, $n_\ye$ as a function of the lateral field $F$ and the gate voltage, $V_\yG$. (c) Saturation velocity versus electron density. 
  (d) Current density versus lateral electric field at various gate voltages at $T=300\, \rm{K}$. \tb{e.} Inset: Gate-dependent resistivity at room temperature, for $F=2\, \rm{mV/\m m}$. 
   \label{fig:Transport}}
\end{figure}

The steady-state solution to Eq.~\eqref{eq:DensityRateEq} reads
\Eq{
n_{\ye/\yh}=\sqrt{ (A_{\rm ZK}/B_R) F^{3/2}+n_\yT^2+(\D n/2)^2}\pm \D n/2,
\label{eq:Concentration}
}
where $\D n=n_\ye-n_\yh$ is the doping and $n_T=\p \n T^2/(24 v_\yF^2)$ is the density of thermal excitations. The doping is controlled by the gate voltage, $V_\yG$, through the relation,  $\D n=C_\yG V_\yG/e$, 
assuming no residual carrier concentration
where $C_\yG$ is the gate capacitance \cite{Meric2008}. Fig.~\ref{fig:Transport}b shows the electron concentration, $n_\ye$, as a function of the gate voltage and the electric field, according to Eq.~\eqref{eq:Concentration}. The density of holes can be found by $n_\yh(V_\yG)=n_\ye(-V_\yG)$.
\vspace{1.2mm}

The current density of the two populations reads $\vJ=\vJ_\ye+\vJ_\yh$, 
where the electronic current is given by $\vJ_\ye=-e \n\int \frac{d^2\vk}{(2\p)^2} \partial_{\vk}E_\yc(\vk) f_\ye(\vk)$ and similarly for the holes. Assuming $\Delta \vk_\ye=\D k_\ye \hat \vx$ and $\D \vk_\yh=\D k_\yh \hat \vx$, where $\D k_\ye$ and $\D k_\yh$ are small, we find
\Eq{
J_{\ye(\yh)}=e n_{\ye(\yh)} v_{\yd,\ye(\yh)}.
\label{eq:CurrentDefinition}
}
Here, $J_{\ye(\yh)}=|\vJ_{\ye(\yh)}|$, $v_{\yd,\ye(\yh)}=v_\yF \D k_{\ye(\yh)}/k_{\ye(\yh)}^\yF$ is the electron (hole) drift velocity and $k_{\ye(\yh)}^\yF=\sqrt{4\p n_{\ye(\yh)}/\n}$ is the ``Fermi-momentum'' of the electron (hole) population. In what follows, for simplicity, we will discuss the electrons in the conduction band, similar considerations apply to holes in the valence band.
\vspace{1.2mm}

In the low field regime, the current of the conduction electrons is given by $J_{\ye}=e n_\ye \m_\ye F$, where $\m_\ye$ is the electron mobility. We note that the mobility controlled by acoustic phonons scales linearly with inverse temperature~\cite{Hwang2008}, at high temperatures, $T\gtrsim 50 \, K$. 
In this regime, the current depends superlinearly on $F$, due to the dependence of $n_\ye$ on $F$, as follows from Eq.~\eqref{eq:Concentration}. 
Comparing with Eq.~\eqref{eq:CurrentDefinition}, this results in $\Delta k_\ye=\Delta k_\ye^0(F)$, where we defined $\Delta k_\ye^0(F)= k_{\ye}^\yF\frac{\m_\ye F}{v_\yF}$.  
\vspace{1.1mm}

At higher electric fields $F> F_{\rm sat}$, the current saturates~\cite{Meric2008} due to two main mechanisms [see Fig.~\ref{fig:Transport}a]. The first mechanism arises when $\Delta k_\ye=k_\ye^F$, where the electrons reach the maximum drift velocity $v_{\yd,\ye}=v_\yF$~\cite{Berdyugin2022}.  The field at which the effect becomes dominant, $F=F_{\rm ZK}$, is given by $F_{\rm ZK}=v_\yF/\m_\ye$, which we estimate by $\sim 0.085\, V/\m m$. 
The second mechanism arises when the electrons reach the energy of optical phonons $\W_{\rm op}$, $\Delta k_\ye=\W_{\rm op}/v_\yF$. To incorporate the saturation mechanisms, we adopt the following model of the saturation~\cite{Yang2017},
\Eq{
\D k_{\ye}\inv=(\D k_\ye^0)\inv+(\D k_\ye^{\rm sat})\inv,
\label{eq:MomentumShift}
}
where $\D k_\ye^{\rm sat}=[(k_\ye^\yF)^2+(\W_{\rm op}/v_\yF)^2]^{-\half}$.
In Fig.~\ref{fig:Transport}c, we plot the drift velocity at saturation as a function of the electron density, defined as $v_{\ye}^{\rm sat}=v_\yF \Delta k_\ye^{\rm sat}/k_\ye^F$. At low electron density, the electroncs can reach the maximal velocity of $v_\yF$, and the second mechanism occurs at stronger fields, $F>F_{\rm ZK}$. however, as the density increases, the velocity saturates because of the optical phonon emission. 



Fig.~\ref{fig:Transport}d, shows the current of the system as a function of the lateral field for various gate voltages at $T=300\, {\rm K}$, and Fig.~\ref{fig:Transport}e shows the resistance of the system as a function of the gate voltage for $F=2\, {\rm mV/\m m}$.

\section{Mid-infrared electroluminescence measurement}

The HPhP electroluminescence is expected to occur at 6-7 $\mu$m and 11-13 $\mu$m. In these wavelength regions, the room temperature thermal emission background can be dominant. To accomplish the measurement of mid-IR emission with suppressed environmental background emission signal, we combined the microscope emission mode of the FTIR and the phase-sensitive detection scheme\cite{zhang2012fourier}. \\

\begin{figure}
	\begin{centering}
	\includegraphics[width=0.9\textwidth]{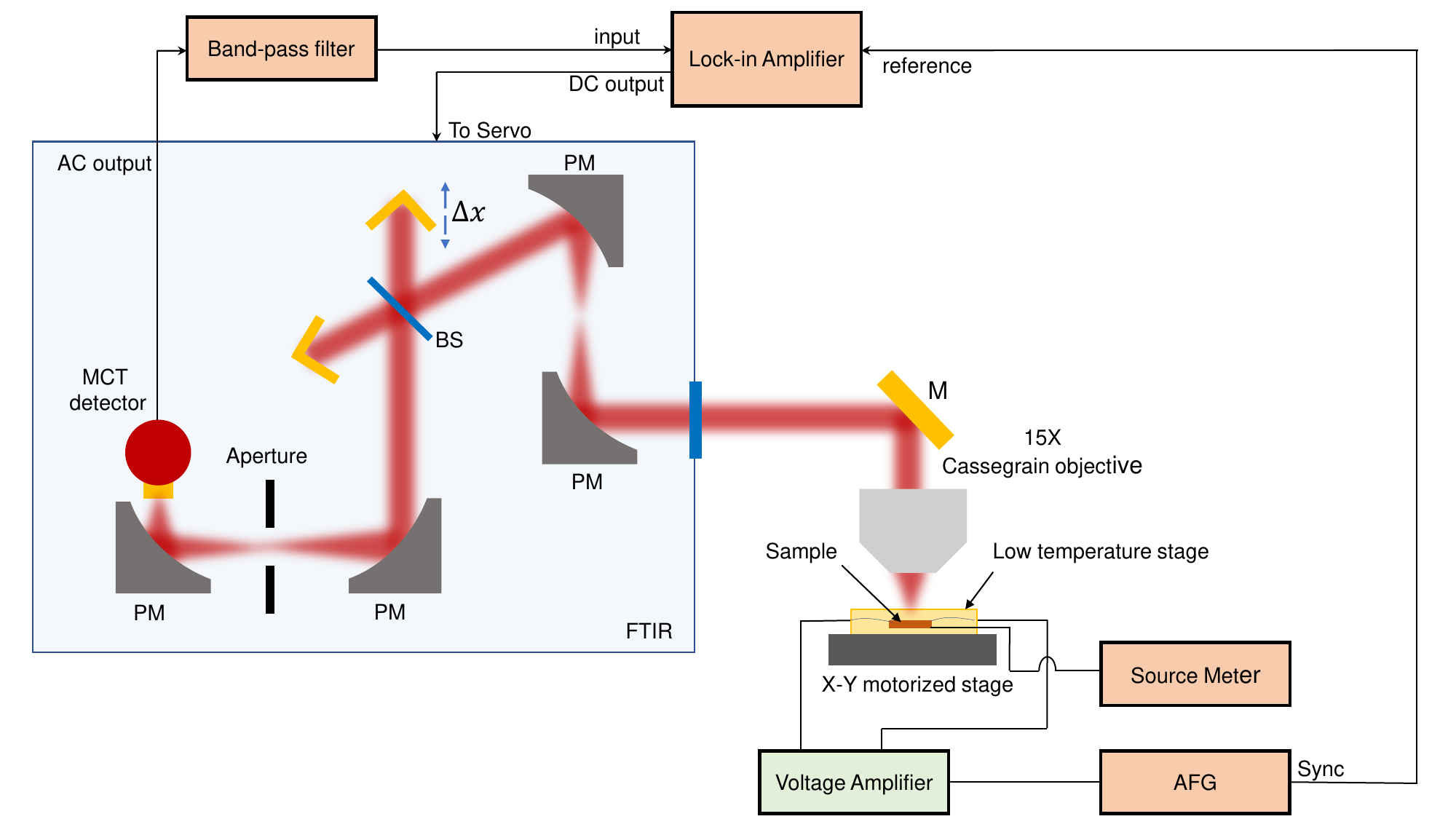}
	\par\end{centering}
	\caption{Schematic diagram of the FTIR spectroscopy setup for mid-IR electroluminescence measurement. BS: beam splitter, PM:parabolic mirror, AFG:arbitrary function generator. }\label{EL}
\end{figure}

As shown in Fig.\,\ref{EL}, the key component in our mid-IR electroluminescence setup is a Bruker Vertex 70 Fourier Transformed Infrared Spectrometer (FTIR) in conjunction with the Hyperion 2000 infrared microscope. The FTIR was operated under the microscope emission mode, which ensures that the infrared radiation from the sample mounted on the microscope stage can be collected and analyzed by the spectrometer and subsequently the photodetector. In our measurement, the sample was wired-bonded to a chip carrier. The sample and the chip carrier were mounted onto a Linkham temperature controllable stage (HFS600-PB4). The gate voltage to the device was controlled by a Keithley 2612 sourcemeter. Instead of using a DC source-drain bias voltage, the source-drain voltage bias was a 34.5 kHz square wave with phase synchronized to the lock-in amplifier. The square wave was provided by an arbitrary function generator (AFG) and subsequently amplified by a homemade voltage amplifier. Once the square wave is applied to the device, the MCT detector collects a DC environmental thermal emission background signal and an AC emission signal from the device. The AC voltage output of the MCT detector was first filtered by a band-pass filter with a narrow pass-band (34.2 - 34.8 kHz) and the filtered output voltage signal was picked up by the lock-in amplifier. The time constant of lock-in amplifier was set to be 1 ms. The DC output signal from the lock-in amplifier was directed to the DC input of the FTIR, which was used to generate the interferograms and the IR emission spectra after performing the Fourier transform. During each FTIR emission spectrum measurement, 32 scans were carried out. To improve the signal-to-noise ratio, we average four obtained spectra at the same measurement condition.

\section{Control experiment}

\begin{figure}
	\begin{centering}
	\includegraphics[width=0.9\textwidth]{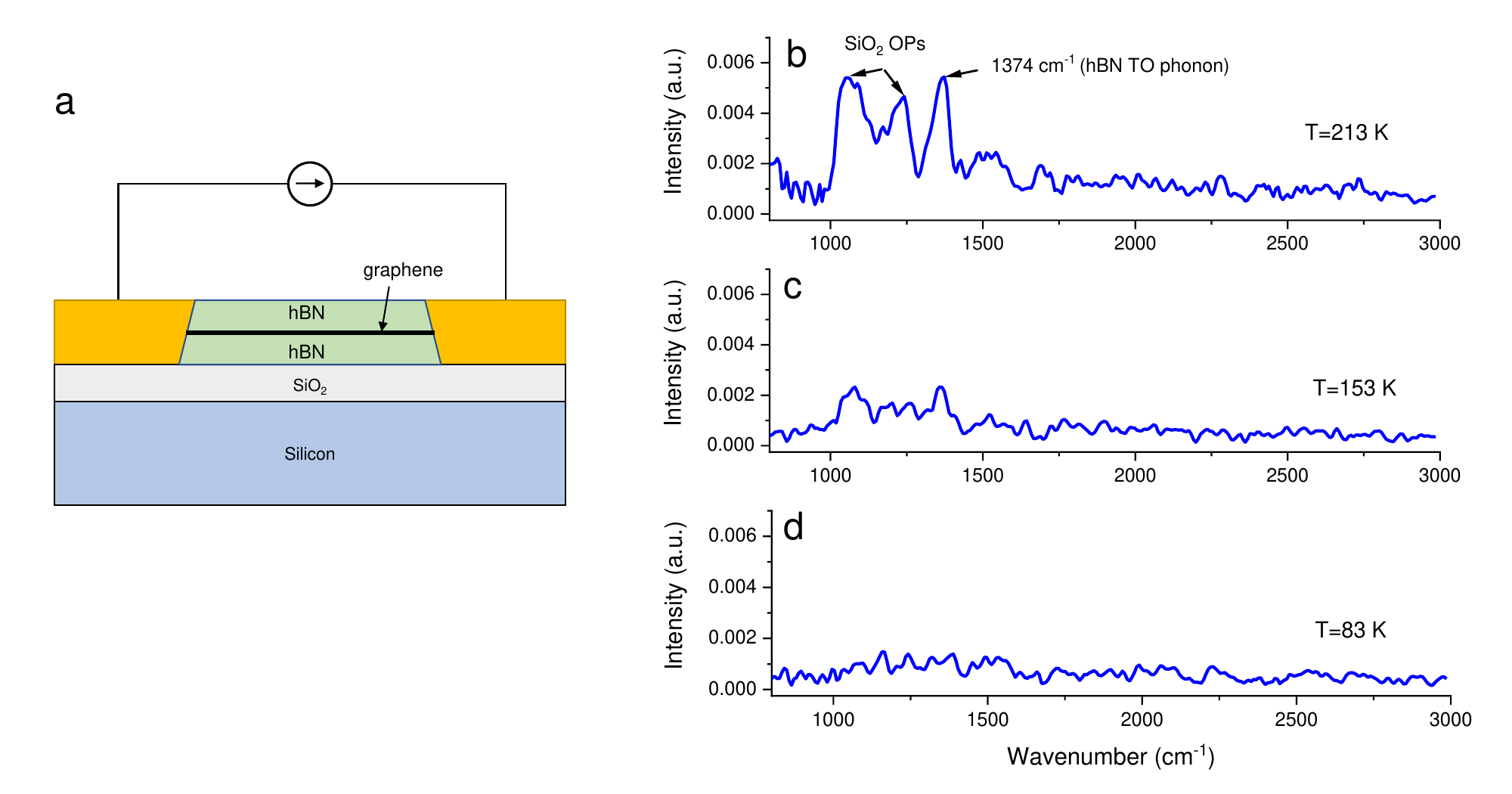}
	\par\end{centering}
	\caption{(a) Illustration of the experimental device used in the control experiment. Mid-infrared emission spectra of hBN/Graphene/hBN sample without Au nanodiscs measured at 213 K (b), 153 K (c), and 80 K (d). }\label{control}
\end{figure}

In addition to HPhP electroluminescence, thermal radiation can also lead to infrared radiation. Here, we performed a control experiment which further confirms that the observed infrared emission is not a result of thermal radiation due to the strong Joule heating of the sample. We performed similar infrared emission measurements on a hBN/Graphene/hBN sample without Au nanodiscs on top as HPhP out-couplers, as shown in Fig. S\ref{control}a. For this sample, we expect thermal radiation will dominate the emission spectrum whereas electrically generated HPPs can be hardly coupled out. The resulting emission spectra at three different environmental temperatures (213 K, 153 K, and 80 K) are plotted in Fig.\ref{control} b, c, and d. When the applied electric power exceeds 500 mW (applied electric field $>$ 2.6 V/$\mu$m), distinct and shark emission peaks corresponding to the TO phonons of hBN (1374 cm$^{-1}$) and SiO$_2$ can be clearly identified. Notably, these peaks are located at different frequencies compared to the emission peaks shown in Fig. 3 and 4 in the main text and there are no obvious emission peaks in the two Reststrahlen bands. According to Kirchhoff’s law of thermal radiation\cite{greffet1998field}, the frequency-dependent emittance of an object is equal to its frequency-dependent absorptance. Thus, the strong infrared absorption of hBN and SiO$_2$ at their respective TO phonon frequencies give rise to the high thermal emissivity at these frequencies. As a unique feature of thermal radiation, the emittance strongly depends on the temperature of the thermal emitter. As shown in Fig. S\ref{control} b to d, the emission peaks were strongly suppressed when the sample was cooled down. Such behavior is drastically different from what we previously observed in Fig. 3f in the main text, in which the emission intensity is almost independent of the temperature. 

\section{Phonon-polariton modes}
\begin{figure}
  \centering
  \includegraphics[width=13.6cm]{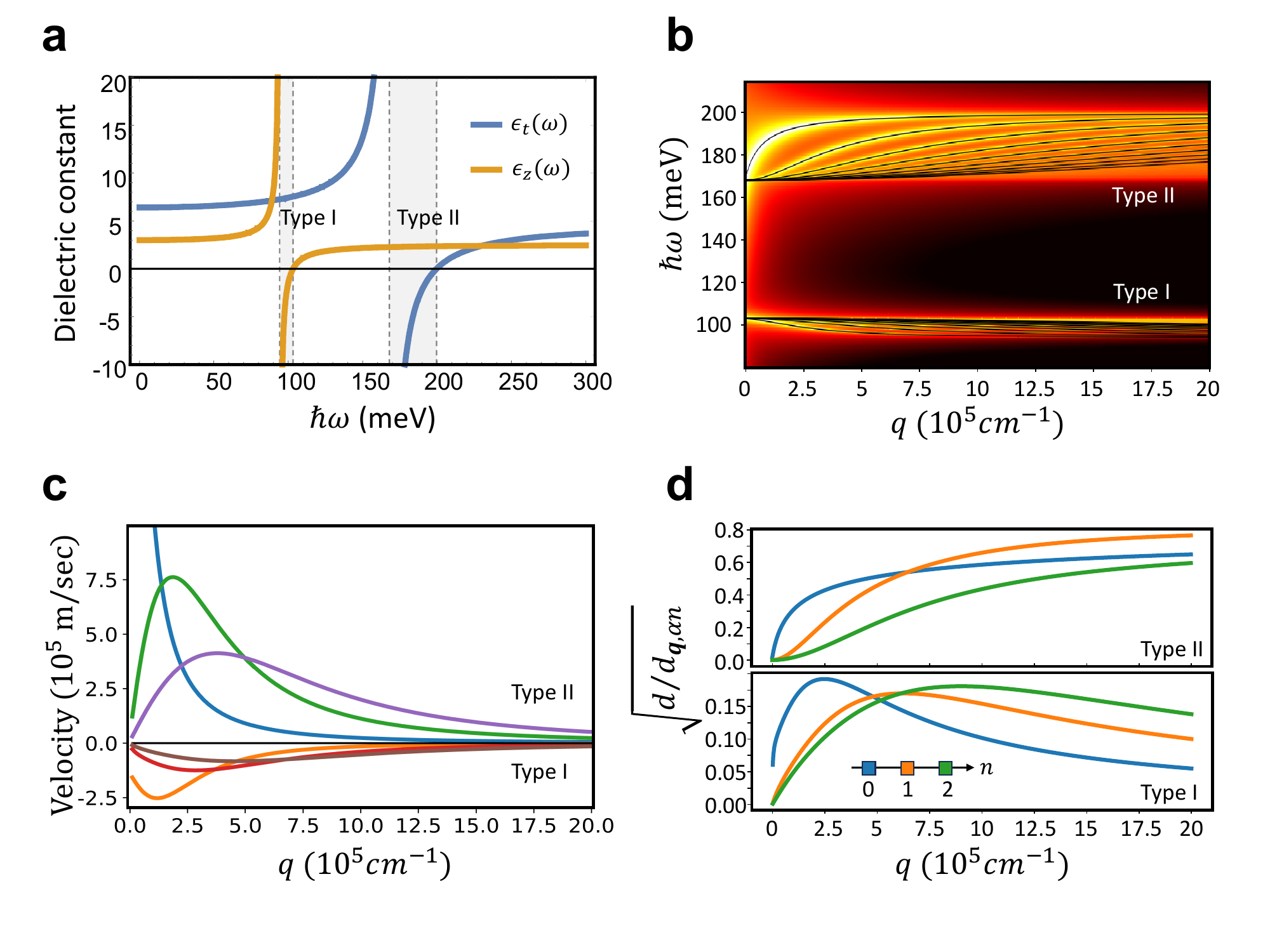}\\
  \caption{(a) Components of the dielectric function of h-BN in the in-plane ($\e_x$) and out-of-plane ($\e_z$) directions. The shaded area denotes frequency ranges where the dielectric functions exhibit opposite signs giving rise to the hyperbolic modes. (b) The imaginary part of the normalized Coulomb potential [given in Eq.~\eqref{eq:RenormalizedCoulomb}] and dispersions of the HPhP modes, $\w_{q,\a n}$ [black lines]. (c) The group velocities of the HPhPs modes.  (d) Effective confinement length. 
   \label{fig:Hyperbolic}}
\end{figure}

We consider the setup, where the electrons coexist with hyperbolic phonon-polariton (HPhP) modes in h-BN. These modes are unique to layered polar insulators, with an anisotropy between in-plane and out-of-plane polar optical phonon modes, giving rise to different dielectric constants $\e_\ell(\w)$ in the in-plane ($\ell=x$) and out-of-plane  ($\ell=z$) directions. The dielectric constants can be parameterized by~\cite{Tomadin2015} 
\Eq{
\e_{\ell}(\w)=\e_{\ell,\infty}+\frac{\e_{\ell,0}-\e_{\ell,\infty}}{1-(\w/\w_\ell)^2-i\g_\ell \hb\w/(\hb\w_\ell)^2},
}
where $\w_\ell$ is the phonon frequency and $\g_\ell$  is the loss rate, see Tab.~\ref{tab:Parameters} for the characteristic values used in the simulation.
Such an anisotropic dielectric function gives rise to two frequency bands in which the signs of $\e_x$ and $\e_z$ are opposite, known as the type I and type II Reststrahlen bands, see Fig.~\ref{fig:Hyperbolic}a. 
The modified dielectric dispersion in a slab geometry gives rise to renormalized Coulomb potential~\cite{Tomadin2015}, 
\begin{equation}
    V_{\vq, \w}=\vf_\vq\frac{\sqrt{\e_x\e_z}+\e_b\tanh(q d \sqrt{\e_x/\e_z})}{\sqrt{\e_x\e_z}+(\e_x\e_z+\e_b\e_a)\tanh(q d\sqrt{\e_x/\e_z})/(2\bar\e)},
    \label{eq:RenormalizedCoulomb}
\end{equation}
where $\vf_\vq=-2\p e/(q\bar \e)$ is the bare Coulomb potential and $\bar \e=(\e_z+\e_b)/2$.

Remarkably, in the Reststrahlen band, the electromagnetic dispersion relation, 
\Eq{
\frac{q^2}{\e_z(\w)}+\frac{\ka^2}{\e_x(\w)}=\frac{\w^2}{c^2},
\label{eq:DispersionRelation}
}
becomes an equation for a hyperbola and therefore has solutions even in the low-frequency limit ($\w^2\ll c^2q^2+c^2\ka^2$), where $q$ is the wavevector amplitude in the in-plane direction and $\ka$ is the out-of plane wave vector. This allows for propagating electromagnetic modes in the bulk of layered materials ~\cite{Dai2014,Dai2015}.\\

In h-BN slab of width $d$, the out-of-plane momentum satisfy the waveguide equation 
\Eq{
\tan(\ka d/2)=-\frac{\ka}{\e_x\sqrt{q^2-\frac{\w^2}{c^2}}}.
\label{eq:WaveguideEq}
}
Solving Eqs.~\eqref{eq:DispersionRelation} and \eqref{eq:WaveguideEq} 
 for $\gamma_\ell=0$ gives rise to an equation for the in-plane momentum of the HPhPs,
\Eq{
q_{\a,n}=\frac{1}{2d}\sqrt{-\frac{\e_z}{\e_x}}\bS{\p(2n+1)\pm 2\tan^{-1}\bR{\frac{\e_a\e_b+\e_x\e_z}{2\bar\e\sqrt{-\e_x\ve_z}}}},
\label{eq:MomentumHPhP}
}
characterized by the integer $n=0,1,2,...$ and the type of Reststrahlen branch $\a=I,II$. 
The dispersion of HPhPs, $\w_{\vq,\a n}$ can be found by inverting Eq.~\eqref{eq:MomentumHPhP}. Fig.~\ref{fig:Hyperbolic}b shows the imaginary part of the normalized Coulomb potential [see Eq.~\eqref{eq:RenormalizedCoulomb}], with black lines denoting the poles $\w_{\vq,\a n}$ given by Eq.~\eqref{eq:MomentumHPhP}. The group velocity of the modes $\vv_{\ys,\vq,\a n}=\dpa_\vq \w_{\vq,\a n}$ is shown in Fig.~\ref{fig:Hyperbolic}c.
The Hamiltonian describing the HPhP modes reads
\Eq{
\hat H_{\rm HPhP}=\sum_{\vq,\a,n}\hb \w_{\vq,\a n} \hat \g_{\vq,\a n}\dg\hat \g_{\vq,\a n},
}
where the operator $\hat \g_{\vq,\a n}$ creates a boson in the  HPhP mode with the energy $\hb\w_{\vq,\a n}$.

\section{The electron-HPhP coupling}
Next we discuss the coupling of HPhPs to the electrons in graphene. The coupling of the electrons to electromagnetic waves is given by 
$\hat H_{\rm ep}=\int \frac{d^2\vq}{(2\p)^2} \hat \vJ_\vq\cdot \hat\vcA_\vq(z=0)$, where $\hat\vJ_\vq$ is the electron current density operator as a function of the in-plane momentum, $\vq$, and $\hat\vcA_\vq$ is the vector potential of the electromagnetic wave. The vector potential in the slab can be expanded in the eigenmodes,  
\Eq{
\hat\vcA_\vq(z=0)=\sum_{n\a}\cA_{\vq,\a n}\bS{\hat \g_{\vq,\a n}+\hat \g\dg_{-\vq,\a n}},
}
where $\cA_{\vq,\a n}\eqa\hat\be_{\vq,\a n}\sqrt{\hb L^2/(2\e_0\w_{\vq,\a n}d_{\vq,\a n})}$, and $d_{\vq,\a n}$ is the effective confinement length~\cite{Ashida2023}, see Fig.~\ref{fig:Hyperbolic}d. Here, we focus on the TM modes, whose polarization vector has a component parallel to $\hat\vJ$, $\be_{\vq,\a n}=\vq/q$. Next, we employ the continuity equation for the electric current and density to arrive at 
\Eq{
\hat H_{\rm ep}=\sum_{\vq,\a,n,\lm} L\inv g_{\vq,\a n}\hat \ro_{\vq,\lm}\bS{\hat \g_{\vq,\a n}+\hat \g\dg_{-\vq,\a n}},
}
where $\hat\rho_{\vq,\lm}=\sum_\vk \hat c_{\vk+\vq,\lm}\dg \hat c_{\vk,\lm}$ is the electronic density operator in the conduction ($\lm=+1$) or valence ($\lm=-1$) bands of graphene, and $g_{\vq,\a n}=\sqrt{\hb 
\w_{\vq,\a n}e^2/(2\e_0 d_{\vq,\a n} q^2)}$ is the electron-HPhP coupling coefficient.

\section{Kinetics of the phonon-polaritons}

Next, we discuss the dynamics of the HPhPs in the system. Here, we derive the kinetic equation from the Keldysh formalism. We then assume quasiparticle approximation, leading to a Boltzmann equation for the HPhPs in the system. \\

First, we define the two-point function of the HPhP modes in the Keldysh representation, $\cD^<_{\a n}(\vr,\vr';t,t')=-i\av{\hat \g\dg_{\a n}(\vr,t) \hat \g_{\a n}(\vr',t')}$ and $\cD^R_{\a n}(\vr,\vr';t,t')=i\av{\com{\hat \g\dg_{\a n}(\vr,t)}{\hat \g_{\a n}(\vr',t')}}\Q(t-t')$, where $\hat\g_{\a n}(\vr,t)=\sum_{\vq \a  n}\hat\g_{\vq,\a n}e^{i\vq\cdot \vr-i\w_{\vq,\a n}t}$, and $\vr=(x,y)$ denotes the position on a 2D surface $z=0$. 
The occupation probability of HPhP modes, $b_{\a n}(\vr,\vr';t,t')$, is then defined by~\cite{Kamenev2011}
$\cD^< =\cD^R\circ b-b \circ [\cD^R]\dg $, where ``$\circ$'' denotes time and space convolutions.\\

The dynamics of $\cD_{\a n}$ is controlled by the free retarded propagator $D^R_{\a n}(\vq,\w)=\frac{1}{\w-\w_{\vq,\a n}+i0^+}$ and the self-energy $\Si_{\a n}(\vq,\w)$ arising from interactions, see Fig.~\ref{fig:Electroluminescence}a. Following Ref.~\cite{Kamenev2011}, we derive the kinetic equation for the Wigner-transformed occupation probability $b_{\vq,\a n}(\vr,t)$, that reads, 
\Eq{
2\dpa_t b_{\vq,\a n}+2\vv_{s,\vq,\a n}\cdot \nabla b_{\vq,\a n}=\Si_{\a n}^K(\vq,\w_{\vq,\a n})-\Si_{\a n}^\D(\vq,\w_{\vq,\a n}) [2b_{\vq,\a n}+1].
\label{eq:KeldyshKineticEquation}
}
Here, $\Sigma^K$ and $\Si^\D$ are the Keldysh and the imaginary part of the retarded component of the self-energy respectively. 
We separate the self-energy to two components, $\Si=\Si_{\rm el}+\Si_{\ta}$, where $\Si_{\rm el}$ arises from the electron-HPhP coupling and $\Si_{\ta}$ arises from other contributions. Using an RPA approximation, the electronic part of the self-energy is proportional to the polarization function of the electrons, $\Si_{\rm el,\a n}=g^2_{\vq,\a n}\sum_{\lm\lm'}\ch_{\lm\lm'}$, where $\ch_{\lm\lm'}(\vq,\w)=\frac{1}{L^2}\int dt e^{i\w t}i\av{\hat\rho_{\vq,\lm}(t)\hat\rho_{-\vq,\lm'}(0)}$ (see the calculation of $\chi$ in Section \ref{section:polarization function}).\\ 

Defining $\Si_{\rm el,\a n}^\D(\vq,\w_{\vq,\a n})=-\G^{\rm amp}_{\vq,\a n}$, $\Si_{\rm el,\a n}^K(\vq,\w_{\vq,\a n})=2\g^{\rm em}_{\vq,\a n}-\G^{\rm amp}_{\vq,\a n}$, $\Si_{\ta}^\D=1/\ta_p$, and $\Si_{\ta}^K=(2b^0-1)/\ta_p$, we arrive at~\cite{Andersen2019}
\EqS{
\dpa_t b_{\vq,\a n}=\G^{\rm amp}_{\vq,\a n} b_{\vq,\a n}+\g^{\rm em}_{\vq,\a n}-\frac{b_{\vq,\a n}-b_{\vq,\a n}^0}{\tau_p}-\vv_{s,\vq,\a n} \cdot \nabla b_{\vq,\a n}.
\label{eq:PhononicKinetic}
}
Here, $\G^{\rm amp}$ is the field amplification rate, $\g^{\rm em}$ is the spontaneous emission rate, $b_{\vq,\a n}^0=[e^{\beta \w_{\vq,\a n}}-1]\inv$ is the equilibrium distribution, and $1/\tau_p=1/\tau_{\rm loss}+1/\tau_{\rm rad}$ is the phonon-loss rate consisting of two main contributions: the loss rate due to phonon-phonon and phonon-impurity interactions, $\tau_{\rm loss}$ and due to the coupling to the outside radiation $\tau_{\rm rad}$. Here, we consider $1/\tau_p=\g_\ell$, where 
The terms proportional to $\G^{\rm amp}$ and $\g^{\rm em}$ arise due to spontaneous and stimulated emission of HPhPs by the electrons.

\section{The electroluminescence of HPhPs}
The steady state ($\dot b=0$) solution to Eq.~\eqref{eq:PhononicKinetic} assuming only changes in the $x$ direction, reads
\Eq{
b_{\vq,\a n}(x)=b_{\vq,\a n}^0+\frac{ \G^{\rm amp}_{\vq,\a n}b^0_{\vq,\a n}+\g^{\rm em}_{\vq,\a n}}{\G^{\rm amp}_{\vq,\a n}-1/\ta_p}[e^{(\G^{\rm amp}_{\vq,\a n}-\frac 1{\ta_p})\frac{x}{v^x_{s,\vq,\a n}}}-1],
\label{eq:HPhPs_density}
}
where $v_{s,\vq,\a n}^x=\hat \vx\cdot \vv_{s,\vq,\a n}$.
The average density of HPhPs per mode is given by integrating the density over the slab area, $L^2$,
$\bar b_{\vq,\a n}=\frac{1}{L^2}\int d^2\vr b_{\vq,\a n}(\vr)$. Explicit integration of Eq.~\eqref{eq:HPhPs_density} yields
\Eq{
\bar b_{\vq,\a n}=B_1 [e^{(\G^{\rm amp}_{\vq,\a n}-\frac{1}{\tau_p})\frac{L}{v^x_{s,\vq,\a n}}}-1]-B_0,
}
where $L$ is the length of the device, $B_0=\frac{b_{\vq,\a n}^0/\ta_p+\g^{\rm em}_{\vq,\a n}}{\G^{\rm amp}_{\vq,\a n}-1/\ta_p}$ and $B_1=\frac{v_s^x(\g^{\rm em}_{\vq,\a n}+b^0_{\vq,\a n}\G^{\rm amp}_{\vq,\a n})}{L(\G^{\rm amp}_{\vq,\a n}-1/\ta_p)^2}$.\\

The intensity of the radiation per mode is given by $I_{\vq,\a n}=\frac{\hb\w_{\vq,\a n}}{\tau_{\rm rad}}\bar b_{\vq,\a n}$, leading to the total intensity per $\w$, 
\Eq{
I_{\rm rad}(\w)=\int \frac{d^2\vq}{(2\pi)^2}\sum_{\a n}I_{\vq,\a n} \dl(\w-\w_{\vq,\a n}).
}

\begin{figure}
  \centering
  \includegraphics[width=12cm]{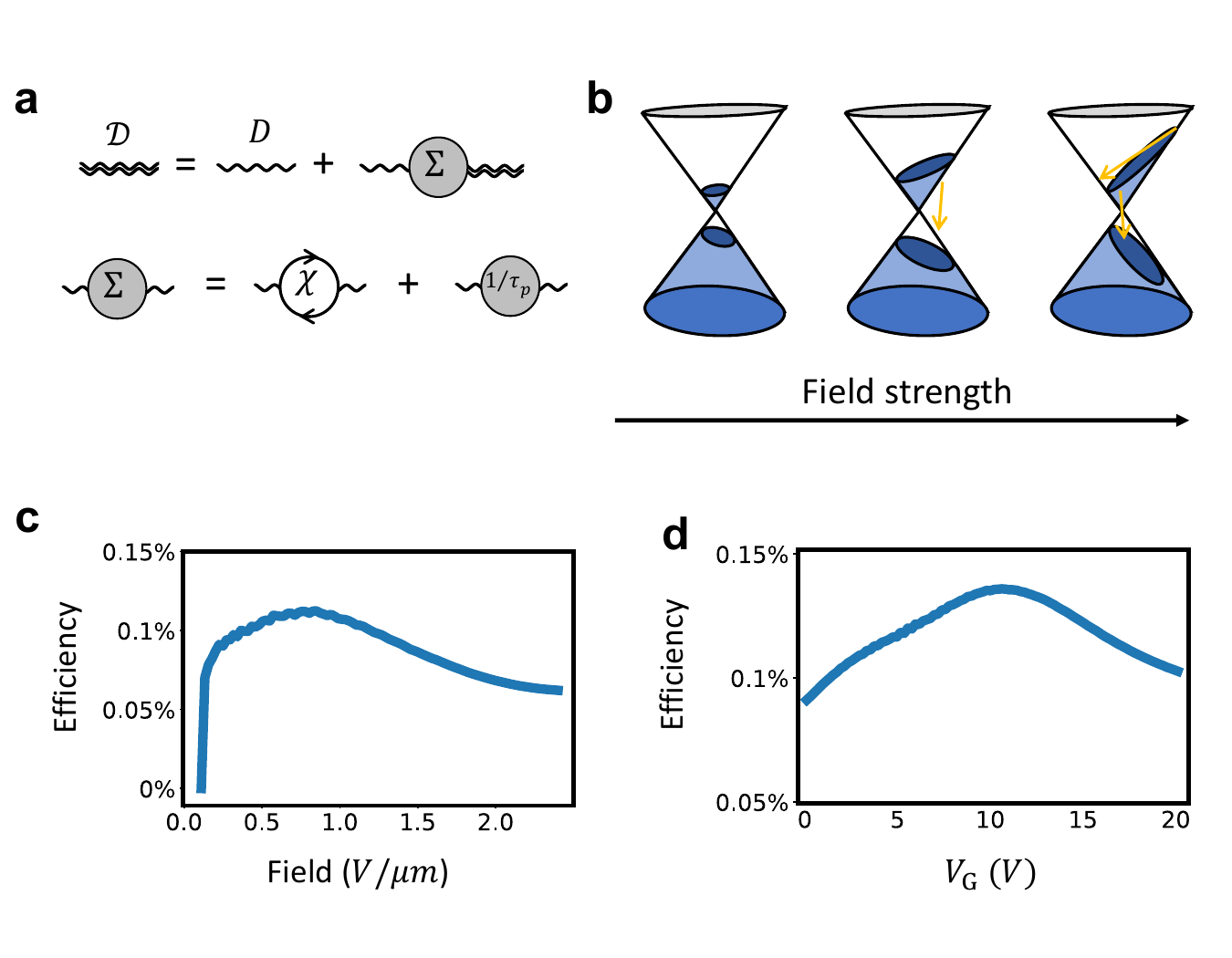}\\
  \caption{ (a) Diagrams leading to HPhP dynamics. (b) Steady state occupation as a function of the electric field. (c,d) Calculated efficiency of the electroluminescence defined as the portion of the emitted intensity from the total dissipated heat (c) as a function of the lateral field strength, and (d) as a function of the gate voltage. 
\label{fig:Electroluminescence}}
\end{figure}

In Fig.~\ref{fig:Electroluminescence}c,d we estimate the efficiency of the HPhP electroluminescence as a function of the field, $F$, and as a function of the gate voltage $V_\yG$. We define the efficiency as the ratio between the total intensity of the HPhP emission $\cI_{\rm tot}=\int d\w I_{\rm rad}(\w)$ and the power dissipated in graphene per area, $\cP=F\cdot J$. We note that the efficiency can be further enhanced if the gain of the device exceeds the total loss, giving rise to a buildup of a strong HPhP field in the device.


\begin{table}
\centering
\begin{tabular}{c| c}
\hline
$v_\yF=10^6 \, {\rm \frac{m}{sec}}$  & $\mu_{\ye,\yh@300K}=117920 \, {\rm \frac{cm^2}{V sec}}$\\
$\W_{\rm op}=200 \, {\rm meV}$  & $A_{\rm ZK}/B_R=2\times 10^{-4}\, {\rm V^{-3/2}\AA^{-5/2}}$\\
\hline
$d=60\, {\rm nm}$& $\e_a=\e_b=1$\\
$\e_{t,0}=6.41$&$\e_{z,0}=3.0$\\
$\e_{t,\infty}=4.54$&$\e_{z,\infty}=2.5$\\
$\g_{t}=0.82\, {\rm meV}$&$\g_{z}=0.23\,{\rm meV}$\\
$\w_{t}=168\, {\rm meV}$&$\w_{z}=94.2\,{\rm meV}$\\
\hline
\end{tabular}
\caption{Parameters used in the numerical simulation}
\label{tab:Parameters}
\end{table}

\section{Calculation of the polarization function}\label{section:polarization function}

Here, we compute the polarization function in the non-equilibrium steady state given by Eq.~\eqref{eq:Occupation} within the RPA approximation. (The polarization function in equilibrium is calculated in Ref.~\cite{Wunsch2006}. ) Following the diagrammatic representation in Fig.~\ref{fig:Electroluminescence}a, we calculate the Keldysh component, $\chi^K_{\lm\lm'}(\vq,\w)$ and $\chi^\D_{\lm\lm'}(\vq,\w)$, defined as $\chi^\D={\rm Im}\chi^R$, reading
\begin{subequations}
\begin{eqnarray}
&&\ch^\D_{\lm\lm'}(\vq,\W)=\frac{\n}{16\pi^3} \int d^2\vk d\w G_\lm^\D(\vk,\w) G_{\lm'}^\D(\vk-\vq,\W-\w)S_{\lm\lm'}(\vk,\vq)[F_{\lm'}(\vk-\vq)-F_{\lm}(\vk)]\\
&&\ch^K_{\lm\lm'}(\vq,\W)=\frac{\n}{16\pi^3} \int d^2\vk d\w G_\lm^\D(\vk,\w) G_{\lm'}^\D(\vk-\vq,\W-\w)S_{\lm\lm'}(\vk,\vq)[F_{\lm}(\vk)F_{\lm'}(\vk-\vq)-1].
\end{eqnarray}
\label{eq:PolarizationIntDef}
\end{subequations}
Here, $G_\lm^\Delta(\vk,\w)=2\p \dl (\w-E_\lm(\vk))$, where $E_\lm(\vk)=\lm v_F |\vk|$ is the dispersion of the upper and lower Dirac cones corresponding to $\lm=\pm$. The function $F_\lm(\vk)=1-2 f_\lm(\vk)$ denotes the non-equilibrium occupation probability with $f_\lm(\vk)\in [0,1]$, and $S_{\lm\lm'}(\vk,\vq)=\half (1+\lm\lm'\frac{k+q\cos(\q)}{|\vk+\vq|})$ is the band-overlap function for graphene, where $\q$ is the angle between $\vk$ and $\vq$ and $k=|\vk|$ and $q=|\vq|$. Throughout, we work in the units of $\hb=1$.\\

Using an explicit expression of $G^\D$ above, one can perform the frequency integral and the integral over $|\vk|$, remaining only with the angular integral
\begin{equation}
\ch^P_{\lm\lm'}(\vq,\W)=\frac{\n}{4\p}\int_{-\p}^{\p} d\q \frac{k_0}{|g'(\vk_0)|}S_{\lm\lm'}(\vk_0,\vq)\cF^P_{\lm\lm'}(\vk_0,\vq).
\label{eq:PolarizationInt1}
\end{equation}
Here, $g(\vk)=\W+\ve_{\lm'}(\vk-\vq)-\ve_{\lm}(\vk)$; the vector $\vk_0$ satisfies $g(\vk_0)=0$, giving rise to an angle-dependent amplitude $k_0(\q)= \frac{v_F^2q^2-\W^2}{2v_F[v_F q\cos(\q)-\lm \W]}$, and $\cF^\D_{\lm\lm'}(\vk,\vq)=F_{\lm'}(\vk-\vq)-F_{\lm}(\vk)$ and $\cF^K_{\lm\lm'}(\vk,\vq)=F_{\lm}(\vk)F_{\lm'}(\vk-\vq)-1$ for $P=\bC{\D,K}$. \\

At zero temperature, $F_\lm(\vk)=-1$ if $E_\lm(\vk-\vk_\lm)<\ve_\lm$ and $F_\lm(\vk)=+1$ otherwise. Here, $\ve_\lm$ is the chemical potential, and $\vk_\lm$ is the population's momentum shift. The integral in Eq.~\eqref{eq:PolarizationInt1} is performed over a region in momentum space where $F_\lm$ and $F_{\lm'}$ yield opposite values. For $F_{\lm'}(\vk-\vq)=+1$ and $F_{\lm}(\vk)=-1$, $\cF^\D=+2$, and $\cF^\D=-2$ otherwise, whereas $\cF^K=-2$ for both cases. \\

Using the explicit forms of  $|g'(k_0)|=\frac{2v_F (v_Fq \cos(\q)-\lm \W)^2}{\W^2-2\lm\W v_Fq \cos(\q)+v_F^2 q^2}$, and $S_{\lm\lm'}(\vk_0,\vq)=\frac{v_F^2 q^2\sin^2(\q)}{\W^2-2\lm \W v_F q \cos(\q)+v_F^2 q^2}$, the integral in Eq.~\eqref{eq:PolarizationInt1} reads
\begin{equation}
\ch^P_{\lm\lm'}(q,\W)=\frac{\n q}{4\p v_F}\int_{\q_1}^{\q_2} d\q \frac{(1-z^2)\sin^2\q}{(\cos\q-\lm z)^3}\cF_{\lm\lm'}^P(\q), 
\label{eq:PolarizationInt2}
\end{equation}
where, we defined $z=\W/(v_Fq)$ and $\q_1,\q_2$ are defined by the range of $\q$ where $\cF^P\ne 0$ . 
The $\q$-integral can be performed exactly yielding
\EqS{
&\ch^P_{\lm\lm'}(q,\W)=\frac{\n q}{4\p v_F}\cF_{\lm\lm'}^P(\q)[\cG(\q_2)-\cG(\q_1)],
}
where 
\Eq{\cG(\q)=\frac{(1-\lm z\cos\q)\sin \q}
{2(\cos\q-\lm z)^2}-\frac{\tanh\inv\bS{\frac{1+\lm z}{\sqrt{1-z^2}}\tan(\frac\q 2)}}{\sqrt{1-z^2}}.
\label{eq:Polarization1}
}

Our next goal is to determine the limits of the integral, $\q_1$ and $\q_2$, that correspond to the points at which $F_\lm(\vk)$ changes sign, i.e., for $E_\lm(\vk_0-\vk_\lm)=\ve_\lm$, yielding an equation for $\q$,
\Eq{
k^2_0(\q)-2 k_0(\q) k_\lm \cos(\q-\f)+k_\lm^2=\ve^2_\lm/v^2_F,
\label{eq:FindBoundary1}
}
where $\f$ is the angle between $\vk_\lm$ and $\vq$, and we assume $|\m_\lm|>2 v_F |\vk_\lm|$.
Similarly, $F_{\lm'}(\vk-\vq)$ changes sign for $\q$ satisfying
\EqS{
k_0^2+q^2+k_{\lm'}^2-2k_0q\cos\q-2k_0k_{\lm'}\cos(\q-\f)+2k_{\lm'}q\cos(\f)
=\ve_{\lm'}^2/v_F^2.
\label{eq:FindBoundary2}
}
These equations can be solved numerically for $\q$. To solve analytically, we assume $\f=0$. 
For this value of $\f$, the system has a symmetry $\q\to -\q$. We, therefore define $x=\cos\q$, assuming $0\le\q\le\p$, for $|x|\le 1$, with double degeneracy for each $x$ (due to $-\p\le\q\le 0$). 
Using this assumption we obtain
\Eq{\cG(x)=\frac{(1-\lm zx)\sqrt{1-x^2}}
{(x-\lm z)^2}-\frac{2\tanh\inv\bS{\frac{1+\lm z}{\sqrt{1-z^2}}\sqrt{\frac{1-x}{1+x}}}}{\sqrt{1-z^2}}.
\label{eq:Polarization2}
}
Here, we multiplied the expression by $2$ to accommodate the double degeneracy. Eq.~\eqref{eq:FindBoundary1} becomes a quadratic equation for $x$ with two solutions
\Eq{
x_1^{\pm}=\lm z+(1-z^2)\frac{\pm\sqrt{z_F^2+k_{\lm}(1-k_{\lm})(1-z^2)}-\lm z k_{\lm}}{2z_F^2-2k_{\lm}(z^2+k_{\lm}-1)},
\label{eq:Boundary1}
}
where $z_F=\m_\lm/(v_Fq)$ and $k_\lm$ is normalized by $q$.
Similarly, Eq.~\eqref{eq:FindBoundary2} can be solved to obtain
\Eq{
x_2^{\pm}=\lm z+(1-z^2)\frac{\pm\sqrt{z_F^2-k_{\lm'}(1+k_{\lm'})(1-z^2)}-
\lm z(1+k_{\lm'})}{2z_F^2-2(z^2+k_{\lm'})(1+k_{\lm'})}.
\label{eq:Boundary2}
}

To find $\ch$, we order all the solutions $(x_1,x_2,x_3,...)$ within the range $(x_{\rm min},1)$, and calculate 
\Eq{
\ch^P_{\lm\lm'}=\frac{\n q}{4\p v_F} \sum_i \cF^P(\bar x_i)[\cG(x_{i+1})-\cG(x_i)], 
}
where $\bar x_i=(x_i+x_{i+1})/2$. The minimal boundary of the range is given by $x_{\rm min}=-1$ if $z>1$, otherwise $x_{\rm min}=\lm z$. Furthermore, if $\lm\lm'=+1$, the result is non-zero for $0\le z\le1$, whereas for $\lm=+1$, and $\lm'=-1$ the result is non-zero for $z\ge1$. We note that for $\lm=-1$, and $\lm'=+1$, the result is always zero. \\

So far we computed $\chi$ for free electrons represented by $G^{\D}$ proportional to a delta function. Finally, we mimic the smearing of the gain function by disorder and finite temperature associated with the timescale $\tau$, by convolving $\chi(\q,\W)$ in energy and momentum, with a broadening function $g(\vq,\W)=\frac{\tau}{\sqrt{2\pi}}e^{-\frac{\W^2\tau^2}{2}}\frac{v_F \tau}{\sqrt{2\pi}}e^{-\frac{v_F^2 q^2\tau^2}{2}}$. This gives rise to a smooth dependence of $\chi$ on the energy and momentum.